# Comparing methods for handling missing data in electronic health records for dynamic risk prediction of central-line associated bloodstream infection

GAO Shan[1], MSc; ALBU Elena[1], MSc; Stijnen Pieter[2], PhD; RADEMAKERS Frank[3], MD; COSSEY Veerle[1,4], MD; DEBAVEYE Yves[5], MD; JANSSENS Christel[6], RN; VAN CALSTER Ben[1,2,7,*], PhD; WYNANTS Laure[1,7,8], PhD

1 Department of Development and Regeneration, KU Leuven, Leuven, Belgium

2 Management Information Reporting Department, University Hospitals Leuven, Leuven, Belgium

3 Faculty of Medicine, KU Leuven, Leuven, Belgium

4 Department of Infection Control and Prevention, University Hospitals Leuven, Leuven, Belgium

5 Department of Cellular and Molecular Medicine, University Hospitals Leuven, Leuven, Belgium

6 Nursing PICC team, University Hospitals Leuven, Leuven, Belgium

7 Unit for Health Technology Assessment Research (LUHTAR), KU Leuven, Leuven, Belgium

8 School for Public Health and Primary Care, Maastricht University, Maastricht, the Netherlands

Corresponding author. Herestraat 49 - box 805, 3000 Leuven, Belgium. Tel: 003216377788; *E-mail address:* ben.vancalster@kuleuven.be

# Abstract

**Background** Electronic health records (EHR) often contain varying levels of missing data. This study aimed to compare different imputation strategies to identify the most appropriate missing data handling approach for predicting central line-associated bloodstream infection (CLABSI) in the presence of competing risks using EHR data.

**Methods** We analyzed data from 30,862 catheter episodes at University Hospitals Leuven from 2012 to 2013 to predict 7-day risk of CLABSI using the landmark cause-specific supermodel, accounting for competing risks of hospital discharge and death. Imputation strategies varied from simple methods (median/mode and last observation carried forward) to advanced techniques such as multiple imputation, regression-based methods, and mixed-effects models that leveraged the longitudinal nature of the data. Random forest imputation which preserves interactions and non-linear relationships was also assessed. We also considered the use of missing indicators combined with all other imputation approaches. Model performance was evaluated dynamically at daily landmarks up to 14 days after catheter placement.

**Results** The missing indicator approach demonstrated the highest discriminative ability, achieving a mean area under the receiver operating characteristic curve (AUROC) of up to 0.782 and superior overall performance based on the scaled Brier score. Combining missing indicators with other methods marginally improved performance over standalone approaches, with the mixed model approach combined with missing indicators achieving the highest AUROC (0.783) at landmark day 4, and the missForestPredict approach combined with missing indicators yielding the best scaled Brier scores at earlier landmarks.

**Conclusions** The missing indicator method showed better performance than other imputation strategies in terms of discrimination, calibration, overall performance, and computational efficiency. This suggests that in EHR data, the presence or absence of information may hold valuable insights for patient risk prediction. On the other hand, the use of missing indicators requires caution, as shifts in EHR data over time can alter missing data patterns, potentially impacting model transportability.

**Trial registration** Clinical trial number: not applicable.

# 1 Background

Missing data problems usually present a major obstacle to developing a valid risk prediction model that minimizes risk of bias and concerns regarding applicability in real-world deployment settings. This challenge is particularly evident when using electronic health record (EHR) data, which often have higher levels of missingness than prospective studies, where data are purposefully collected and rigorously cleaned [1]. In EHR, the presence and timing of measurements may vary considerably from patient to patient. The percentage of patients with missing values for certain measurements can exceed 90% [2]. Missing values can arise due to various reasons that are associated with distinct statistical missingness mechanisms [3]. For example, missing completely at random (MCAR) occurs when the likelihood of missing data is unrelated to observed or unobserved factors, such as a measurement being omitted due to an unexpected technical issue with the equipment [4]. Missing at random (MAR) refers to the situations where the probability of missing data is systematically related to observed variables, but not to the unobserved values. For example, Glasgow Coma Scale (GCS) scores may be more frequently recorded in the intensive care unit (ICU) than in other wards, leading to missingness associated with observed factors [4]. On the other hand, missing not at random (MNAR) arises when the missingness depends on unobserved factors, such as patients declining self-reported pain assessments due to severe pain [4]. Different imputation methods make different assumptions about the missing data mechanism and may therefore vary in their performance when applied to datasets with diverse types of missingness. In EHR data, complete case analysis is often problematic because it can introduce bias due to the non-random nature of missingness and also reduce the sample size. When applying prediction models in real time to a new patient based on EHR data, missing values should also be handled in real time. Therefore, imputation should be performed before applying the prediction model, using the imputation model that is already available at prediction time. While including the outcome in the imputation model is commonly recommended for causal research questions, it is not appropriate in prediction settings where the outcome is unknown at the time of prediction [5].

To investigate various imputation methods in the context of central line-associated bloodstream infection (CLABSI) prediction, we utilized the University Hospitals Leuven (UZ Leuven) EHR data as a case study. CLABSI is defined as any laboratory-confirmed bloodstream infection (LCBI) in a patient with a central line or within 48 hours after its removal, excluding cases present on admission, secondary infection, and mucosal barrier injury LCBI [6]. As one of the common sources of hospital-acquired infections, CLABSI poses a significant risk for patients with central venous catheters, often resulting in prolonged hospital stays, increased healthcare cost, and higher patient morbidity [7,8,9]. A valid prediction model can enable hospital staff, such as nurses, physicians and vascular access specialty teams, to identify high-risk patients and implement timely preventive interventions to reduce CLABSI risk. However, missing data problems present a challenge when developing a practically applicable CLABSI risk prediction model [10]. A recent systematic review of published CLABSI prediction models reported how missing data were handled in these studies [11]. From the 16 developed models that were reviewed, 11 did not provide information on the amount of missing values, nor did they report the methods used to handle missing data. Of the models that did address missingness, three relied on complete case analysis, with only one model using multiple imputation and another using missing indicator methods. Furthermore, none of the models discussed the challenges of handling missing data during practical deployment, such as the applicability of imputation using the data available at prediction time when a single new individual presents with missing values.

In this study, we used electronic health records data from UZ Leuven to predict the 7-day risk of CLABSI for hospitalized patients with central venous catheters. We aimed to compare commonly used approaches for addressing missing data challenges in the UZ Leuven EHR data, with an emphasis on optimizing the performance of prediction models for CLABSI. We focused on imputation methods that are built on training data and applicable to a single new observation at prediction time, ensuring their clinical applicability.

## 2 Methods

### 2.1 Study design and participants

The retrospective cohort study includes 27,478 patient admissions from the University Hospitals Leuven where patients received any type of central venous catheter between January 1st, 2012 and December 31st, 2013. These admissions resulted in 30,862 patient-catheter episodes (i.e. a period during which a patient has a catheter), as patients could have multiple catheters being placed during their hospital admission. Consecutive catheters are considered a single patient-catheter episode if the gap between the last observation of one catheter and the first observation of the next is 48 hours or less, otherwise, they are counted as separate catheter episodes. Details can be found in Supplementary File 1.

Patients in the neonatology department were documented using a paper-based workflow before October 2013 and did not have their electronic registries in the system. Thus hospital admissions for patients under the age of 12 weeks have been excluded from the study.

### 2.2 Study outcome

The outcome to be predicted is the occurrence of CLABSI per patient-catheter episode, updated every 24 hours starting from the first recorded catheter observation during the hospital stay. Each time, we estimated the risk of CLABSI within the next 7 days (i.e. the prediction horizon) as suggested by clinical experts to allow timely intervention. CLABSI is defined according to the 2019 guidelines of the Belgian public health institute Sciensano [6]. Additional details are provided in Supplementary File 1. Furthermore, we identified two competing events that preclude the occurrence of CLABSI: (1) death or start of palliative care, and (2) hospital discharge or catheter removal for more than 48 hours without CLABSI. There is no actual censoring in the UZ Leuven EHR data, as each patient was observed with one of the events mentioned above.

In the UZ Leuven EHR data, there were overall 30,862 catheter-episodes, of which 970 (3.1%) resulted in CLABSI. 404 CLABSI events occurred within the first 7 days after the first catheter observation recorded for the catheter episode, 566 CLABSI events occurred later.

## 2.3 EHR data and predictors

EHR data were extracted from various electronic sources including demographics information, patient admissions/discharges, medical specialties, catheter-related observations, patient medication prescriptions, comorbidities, laboratory tests and vital signs. We used 21 predictors in this study (Table 1), which were chosen based on LASSO (least absolute shrinkage and selection operator) selection. Details regarding the variable selection can be found in Supplementary File 2.

## 2.4 Statistical analysis

We developed and validated our models using a repeated data-splitting approach. For each iteration, two-thirds of hospital admissions from the dataset were randomly assigned to training, while the remaining one-third served as the validation set. To avoid potential leakage, all landmarks and catheter episodes of each admission were consistently assigned to either the training or validation set, ensuring no overlap in admission data between sets. The models were trained on the imputed training data and then applied to the validation data, where missing values were imputed using the same imputation model. Model performance metrics were then assessed on the imputed validation data. This process was repeated 100 times and performance metrics were summarized across the 100 train-validation splits.

### 2.4.1 Missing data imputation

The key strategies compared in this study for handling missing data are outlined in Table 2 and detailed further in Supplementary File 3. Median or mode imputation replaces missing values with the median (for continuous variables) or mode (for categorical variables) of the observed data, without considering the temporal structure. Last Observation Carried Forward (LOCF) leverages the temporal nature of the data by imputing missing values based on the most recently observed value within the same catheter episode. While these methods are easy to implement, they may introduce biases, especially in datasets where the missingness is not random [16,17].

Multiple imputation handles missing data by generating multiple imputed datasets to incorporate random variations to reflect the uncertainty in the missing values [18]. In our study, we applied the MICE algorithm to generate ten imputed datasets, using ten iterations per dataset to ensure convergence of the imputation process. At prediction time, each validation dataset was imputed through a new ten-iteration process, initialized with the corresponding imputation model learned from the training data (i.e., the state of the imputer after the last iteration of training). This process produced ten imputed validation datasets, consistent with the number of imputed training datasets. The prediction model, trained on each corresponding imputed training dataset, was then applied to the respective imputed validation dataset. Finally, the resulting predictions were combined across the ten datasets using Rubin's rules to produce a single, pooled prediction for each individual, following the "pooled prediction" strategy [21]. These pooled predictions were then used to calculate the overall performance metrics.

Although multiple imputation can be based on regression models, we use this term in this study to refer specifically to approaches that generate multiple imputed datasets through stochastic sampling. In contrast, we use regression imputation (RI) to describe single imputation methods that create one completed dataset by fitting regression models using the specified method [10]. As a more pragmatic alternative to MI, we also evaluated RI, which imputes missing values in a single step without accounting for imputation uncertainty. At prediction time, RI was applied by imputing missing values in the validation set using the trained imputation model (after one iteration) and then applying the trained prediction model.

It is generally recommended to include the outcome variable in the imputation model when applying multiple imputation to avoid bias in coefficients estimates [5]. While this may be statistically ideal during model development, it is not clinically feasible in prediction settings where the outcome is unknown at the time of prediction. To ensure that predictive performance in model validation aligns with real-world settings, it is appropriate to use the same missing data handling strategy as during development, which makes it reasonable to exclude the outcome from the imputation model during development. Although this may introduce bias in coefficient estimates, it provides a more realistic estimate of predictive performance, reflecting what might be expected during implementation. In addition, unbiased estimates do not necessarily guarantee better predictive performance, especially in real-world settings [22]. In this study, we applied two approaches to multiple imputation either with or without outcome in the imputation model during development: (1) MICE-xx: excluding the outcome variable from the imputation models, which were fit on the training data and directly applied to the validation set; and (2) MICE-yx: including the outcome in the imputation models fitted on the training data, and setting the outcome to missing in the validation to align with real-world deployment scenario where the outcome is unknown at prediction time. When performing imputation on the validation data, MICE treats the outcome as just another incomplete variable and temporarily imputes it using the observed predictors and its learned relationships, despite it being fully observed in the training set. These imputed outcome values were not used for model evaluation, instead, they served only to help impute missing values in the other predictors, leveraging any relationships between the outcome and the predictors. After imputation was complete, the true observed outcome values n the validation data were then restored for evaluating predictive performance. This strategy allows us to assess the impact of including the outcome variable in the imputation model on predictive performance while reflecting both statistical considerations and real-world feasibility.

The mixed model approach involves fitting mixed-effects models using past landmark measurements, inspired by the 2-stage dynamic landmark model [14]. This requires all individuals with at least one measurement of the time-varying predictors. For predictors missing at baseline, we impute their baseline values using the median or mode of the available baseline data. Subsequently, the mixed-effects models were fitted on the training data, with the models saved and used to impute missing values in the validation set. Further details on the model specifications can be found in Supplementary File 3.

The missForestPredict approach addresses missing data by iteratively using random forest-based imputation models, which help preserve interactions and non-linear relationships between variables [15,23]. Imputation is performed by training the random forest imputation models on the training data, and then applying the saved models to predict missing values in the validation set. Also, the outcome variable is excluded from the imputation model to mimic prediction model use in real-world scenarios.

Additionally, the simple method of missing indicators, where the presence of missingness is flagged as a separate variable, was also considered to account for potential information in the missingness itself [24]. Especially EHR data are often not missing at random, as the presence or absence of information on medication prescriptions, laboratory tests, and clinical actions is typically influenced by patient needs and clinical decisions [25]. The missingness itself may carry valuable informative about a patient's condition. Incorporating the information about the absence of patient data in a prediction model may improve the predictive performance [26]. In this study, we considered both the inclusion of missing indicators as predictors in the risk prediction model, with all missing values imputed by fixed values, and combining it with other imputation methods that incorporate their original imputation strategies for missing entries while also including missing indicators as additional variables.

The imputation approaches consider both binary covariates (for example, medical specialty) and continuous covariates (such as, the maximal value of temperature per landmark timepoint and the last pH value since previous landmark timepoint). Laboratory tests were log-transformed prior to imputation for regression imputation, multiple imputation and mixed model imputation approaches. The distributions of laboratory tests before and after log-transformation (if applied) were shown in Supplementary File 4.

*2.4.2 Model fitting*
We fitted a landmark cause-specific model in this study considering its flexibility to calculate cause-specific cumulative incidences within prediction time horizon and its good prediction performance in a previous study [27]. Predictions were updated dynamically at each landmark time point, with a 7-day prediction time horizon. The last prediction was made at landmark day 30, utilizing data available up to that point and assessing outcomes up to day 37, given that only 872 catheter episodes (2.8%) remained at risk beyond day 30. Further details on the landmark approach can be found in Supplementary File 5.

For a dynamic model, smooth baseline hazards were assumed, by including the linear and quadratic landmark time variables over the stacked landmark datasets in model-fitting. As in the previous study, we included the interaction term between ICU presence and landmark time

(both linear and quadratic) in our dynamic models to capture the time-dependent effect of this predictor [27].

*2.4.3 Model evaluation*
We evaluated model performance on validation data by assessing discrimination, calibration, and overall performance [28,29]. Discrimination was measured using the area under the receiver operating characteristic curve (AUROC), which assesses the model's ability to differentiate between patients with and patients without the outcome event of CLABSI [28]. Calibration was assessed by comparing how well the estimated probabilities match the observed probabilities using both visual and quantitative methods. Calibration plots were used to visually assess agreement by plotting observed outcome frequencies against predicted risks, either across deciles or with smoothing curves. Quantitative metrics included the calibration slope (which measures whether predicted risks are systematically over- or underestimated; target value is 1), the observed-to-expected (O/E) ratio (which compares the total number of observed events to the total number of predicted events; a ratio of 1 indicates perfect agreement) and estimated calibration index (ECI) (a summary measure reflecting the average squared difference between predicted and observed probabilities across all individuals, with lower values indicating better calibration) [29,30,31,32]. Overall performance, which measures the distance between the predicted probability and actual outcomes, was evaluated using the scaled Brier score [28]. Given the absence of censoring in the dataset, all performance measures were calculated by treating the outcome as binary. Specifically, we compared the predicted probability of CLABSI within 7 days against the actual occurrence of CLABSI within 7 days (yes vs no).

*2.4.4 Sample size and software*
The sample size calculation was detailed in Supplementary File 6 using the pmsampsize package [33]. All analyses were performed using R v4.4.1. Software packages used for imputing datasets were shown also in Table 2. The codes were illustrated using the UZ Leuven EHR data. Details regarding the R code for data imputation, model fitting and evaluation can be accessed via https://github.com/chendcw/CLABSI_missing_data. As it is not permitted to share the original data, a manually created example dataset is shared in Supplementary File 5 with sensitive information being replaced.

# 3 Results

## 3.1 Comparison of imputation models

As performance results were extremely uncertain at later landmark time points due to the very limited number of CLABSI events, we presented results only up to landmark day 14. For overall performance, the missing indicator method always showed the highest scaled Brier score across all landmark time points (Figure 1). Among the other imputation methods, no obvious differences in their scaled Brier scores were observed before landmark day 8. After landmark day 8, regression imputation had the lowest scaled Brier scores, except at landmark day 12, where LOCF showed slightly worse overall performance.

Discrimination performance increased from landmark day 0 to landmark day 5, then exhibited a declining trend, with the exception of an increase at landmark day 10 (Figure 1). The missing indicator approach showed the highest discriminative ability (mean AUROC up to 0.782), followed by median/mode imputation (mean AUROC up to 0.776). Regression imputation consistently performed the worst across most landmark time points (mean AUROC up to 0.768). No obvious differences in discrimination were observed between MICE-xx and MICE-yx.

Calibration plots for all models across landmarks were presented in Supplementary File 7. Overall, calibration deteriorated over time, potentially due to the substantial decrease in sample size at later landmarks. Calibration slopes up to landmark day 14 were also shown in Figure 1, with MICE-xx and MICE-yx following similar trends. Regarding the O/E ratio, the missForestPredict approach had the O/E ratio closest to 1 at landmark day 0. The median/mode imputation method displayed the O/E ratio closest to 1 from landmark days 5 to 12, except at landmark days 7 and 8, where the missForestPredict approach performed best in terms of O/E ratio. Also, ECI values increased with later landmark days, further indicating worsening calibration over time.

MICE-xx and MICE-yx did not demonstrate much differences in terms of discrimination, calibration, and overall performance. We also compared these with another ideal but unrealistic deployment scenario for MICE, where the outcome is included in the imputation model fitted on the training data, and applied directly to the test data where the outcome is known, call MICE-yy. The results of this comparison were presented in Supplementary File 8, with no obvious differences observed among the three MICE approaches.

The OOB NMSE from missForestPredict provide an estimation of the performance of each variable's imputation (Supplementary File 3). An NMSE close to 1 indicates weak imputation, similar to mean imputation, while an NMSE close to 0 indicates strong imputation. Despite the fact that NMSE values are mostly under 0.5, indicating sufficient information in other variables for imputation, the differences between model-based imputation methods were rather small. This indicates that the relationships between variables are relatively simple, and the added value of random forest imputation, which can exploit non-linearities and interactions, was minimal.

### 3.2 Combining missing indicators with imputation methods

Recognizing that missing indicators may be informative especially in EHR data [26], we additionally applied the imputation methods outlined in Table 2 in combination with missing indicators. In general, all imputation methods that incorporated missing indicators showed better performances than using imputation methods alone. However, the improvements were minimal when compared to the missing indicator approach alone.

Figure 2 showed the comparison of performance metrics of the imputation methods in Table 2 when incorporating missing indicators. The differences in performance across various imputation approaches became smaller after incorporation with missing indicators. For overall performance, the missForestPredict approach with missing indicators had the highest scaled Brier score at the first three landmark time points (Figure 2). Regression imputation with missing indicators still performed worse than other approaches in most landmarks.

In terms of discrimination, the mixed model approach with missing indicators reached the highest AUROC of 0.783 at landmark day 4 (Figure 2). Regression imputation still had the lowest AUROC at most landmarks, after the incorporation of missing indicators.

Calibration plots for all imputation approaches incorporating missing indicators across landmarks were presented in Supplementary File 7. The calibration slopes for MICE-xx and MICE-yx with missing indicators are always higher than other imputation methods (Figure 2). At landmark day 0, the missForestPredict approach with missing indicators demonstrated the best calibration performance, as reflected by its O/E ratio, calibration slope and ECI.

We also compared the performance of MICE-xx and MICE-yx with missing indicators incorporated, alongside the ideal MICE-yy approach, which also incorporated missing indicators. Similar to the comparison without indicators, no notable differences were observed among the three, although all showed improved performance compared to their counterparts without missing indicators.

The OOB NMSE from missForestPredict incorporating missing indicators were also similar to those without missing indicators (Supplementary File 3), suggesting that the inclusion of missingness indicators in the random forest imputation models did not substantially alter imputation accuracy.

## 4 Discussion

In this study, we compared different imputation methods which are feasible for real-world deployment to predict the risk of CLABSI within 7 days during hospital stay using EHR data from UZ Leuven. To our knowledge, this is the first study to compare imputation methods in the context of dynamic prediction for realistic deployment settings using EHRs. Among these methods, the missing indicator approach showed generally better performance than the other imputation strategies in terms of discriminative ability, calibration and overall performance, while also requiring the least imputation time and acceptable runtime for model development and prediction (Supplementary File 9). This finding highlights the valuable information that can be derived from the presence or absence of patient data in EHR data [26].

Unlike prospective cohort studies where investigators can exert control over when and how measurements are taken by preparing and implementing a research protocol, EHR data is typically recorded following changes in a patient's condition or actions taken by clinicians when deemed necessary. Missing data in EHR can result from two main scenarios: (1) records are missing due to unknown reasons, which may require complex tracking through clinical notes and ordering systems to detect the reason (for example, a lab test may have been ordered, but its result is missing from the system); and (2) records are missing because the corresponding measurements were deemed unnecessary. Examples of this include a lab test that was never ordered or routine checks by nurses, where no adverse findings were observed, resulting in no additional records [34]. In the latter case, patients with more missing records may be associated with better health outcomes, just like critically ill patients in the ICU typically have more frequent monitoring and a larger volume of recorded data [35]. Therefore, we consider this type of missingness to be informative. A study found that the timing of the test request for many laboratory measurements was a better predictor of 3-year mortality risk than the actual test result [36]. This aligns with our findings, suggesting that the presence of missing data can be a valuable indicator of health outcomes.

However, the use of missing indicators in prediction models requires caution, as it assumes a consistent missingness mechanism throughout the model pipeline, which may not hold in EHR data, especially over long time periods [37]. Changes in clinical practices or policies, reflected in EHR data, can alter the relationship between the missing indicators and the outcome, potentially impacting model performance. For instance, if nurses are mandated to check a box for routine patient checks as part of their job performance evaluations, the missing data mechanisms may shift, rendering previously informative patterns uninformative. Regular updating or recalibration of prediction models can help accommodate these evolving patterns in EHR data over time, however, such updates should not be reflexive but rather guided by careful evaluation of whether and how the causal mechanisms have changed [38]. Therefore, deploying such models requires careful consideration of whether clinical decision-making might adapt to the model's characteristics, potentially shifting the missing data mechanisms. Future research should explore how changes in missing data mechanisms affect the transportability of prediction models [39,40,41].

In the context of multiple imputation, although we focused on realistic scenarios where the outcome is unavailable at prediction time (MICE-xx and MICE-yx), we also compared these with the ideal but impractical MICE-yy approach. All three methods performed similarly, regardless of whether missing indicators were incorporated. Given that the outcome cannot be predicted with a high predictive performance based on the predictors (scaled Brier score < 0.04 at all landmarks), we can assume the reverse is also true: other variables cannot be predicted with high predictive performance based on the information contained in the outcome. The inclusion of the outcome in imputation methods in settings with higher predictive performance (e.g., scaled Brier score > 0.5) might yield more pronounced differences.

Our study also has limitations. We applied predictive mean matching in the MICE approaches for imputing missing values in the ordered categorical variables, like the GCS total score. Ordinal logistic regression (via "polr" method) may be a more appropriate method for ordinal predictors. However, due to the significant computational time required by these methods, we opted for the "pmm" method, which we considered a reasonable alternative given that the

GCS total score had many levels [13]. Additionally, the findings of this study are limited by the use of EHR data from UZ Leuven covering the period from 2012 to 2013, which may restrict the generalizability of the results. The nature of missing mechanisms can differ considerably across various clinical environments. For EHR data, the most suitable missing data handling strategy may vary between hospitals due to differences in for example population characteristics, healthcare practices, and technical machines. It is also important to ensure that the missing data handling strategy used during model development remains applicable and appropriate at deployment time, given the potential differences in missingness patterns. In this study, our evaluation focused on internal validation using train/validation splits. Notably, LASSO variable selection was also performed on these same data splits, which may introduce optimism in performance estimates due to potential information leakage. However, as the goal of this study was to compare imputation strategies and determine the most suitable missing data handling approach for CLABSI prediction using UZ Leuven EHR data, we consider the impact to be limited. Moreover, model performance on future unseen data is more relevant for clinical applicability. For example, ciclosporin, one of the predictors, was found to be rarely ordered after 2017, leading to a high rate of missingness from that time onwards. Further research could focus on studying the impact of missing indicator approach in EHR data with underlying temporal shift, as evolving hospital processes and recording practices may affect missingness patterns more than the actual distribution of variables. Lastly, while EHR data contains a vast number of variables, we only incorporated those used in the prediction model for imputations. Expanding imputation models to include auxiliary predictors may help further improve the quality of imputations and model performance [42].

## 5 Conclusion

In our study settings, neither simple imputation methods (e.g., median/mode and LOCF) nor more advanced methods, such as MICE, the mixed model approach, and missForestPredict, improved predictive performance compared to the missing indicator approach. The missing indicator method showed superiority over other imputation methods in terms of discrimination, calibration and overall performance, as well as fast computation time. This is in line with earlier research suggesting that the presence or absence of patient data itself can be informative for patient risk prediction. Missing indicator methods should be considered for prediction modeling, however with caution due to potential differences in missing mechanisms between hospitals and within the same hospital across time.

# List of abbreviations

| | |
|---|---|
| AUROC | area under the receiver operating characteristic curve |
| CICC | centrally inserted central catheter |
| CLABSI | central line-associated bloodstream infection |
| ECI | estimated calibration index |
| EHR | electronic health records |
| GCP | good clinical practice |
| GCS | Glasgow Coma Scale |
| GDPR | General Data Protection Regulation |
| ICU | intensive care unit |
| IQR | interquartile range |
| IRB | institutional review board |
| LASSO | least absolute shrinkage and selection operator |
| LCBI | laboratory-confirmed bloodstream infection |
| LM | landmark |
| LOCF | last observation carried forward |
| MAR | missing at random |
| MCAR | missing completely at random |
| MICE | multiple imputation by chained equations |
| MNAR | missing not at random |
| NMSE | normalized mean square errors |
| O/E | observed-to-expected |
| OOB | out-of-bag |
| PICC | peripherally inserted central catheter |
| RI | regression imputation |
| tc-CICC | tunneled cuffed-centrally inserted central catheter |
| TIVAD | totally implanted vascular access devices |
| TPN | total parenteral nutrition |

## Declarations

### Ethics approval and consent to participate

The study adhered to the principles of the Declaration of Helsinki (current version), the principles of Good Clinical Practice (GCP), and all relevant regulatory requirements. The study was approved by the Ethics Committee Research UZ / KU Leuven (EC Research, https://admin.kuleuven.be/raden/en/ethics-committee-research-uz-kuleuven) on 19 January 2022 (S60891). The Ethics Committee Research UZ / KU Leuven waived the need to obtain informed consent from participants. All patient identifiers were coded using the pseudo-identifier in the data warehouse by the Management Information Reporting Department of UZ Leuven, according to the General Data Protection Regulation (GDPR).

### Consent for publication

Not applicable.

### Availability of data and materials

The data underlying this article cannot be shared publicly due to privacy of individuals that participated in the study. Data are located in controlled access database at UZ Leuven.

### Competing interests

The authors declare that they have no conflicts of interests to disclose.

### Funding

This work was supported by the Internal Funds KU Leuven [grant C24M/20/064]. The funding sources had no role in the conception, design, data collection, analysis, or reporting of this study.

### Authors' contributions

SG, EA, BVC and LW conceptualized the study. PS, BVC and LW secured the funding and obtained institutional review board (IRB) approvals. PS managed data collection. FR, VC, YD and CJ provided clinical inputs. SG and EA performed the data preprocessing and analysis. BVC and LW provided methodological guidance on the modeling framework. SG prepared the original draft and all authors participated in revising and editing the article.

### Acknowledgements

Not applicable.

Table 1. Descriptive statistics for the 21 predictors based on 30862 catheter observations including 241501 landmark (LM) observations.

| Predictor | Scale | Median (IQR) or N (%) | N (%) of catheter episodes with missingness |
|---|---|---|---|
| Static predictor [a] | | | |
|     Age at admission (years) | Continuous | 60 (47, 70) | 0 (0%) |
|     Admitted from home | Binary | 26637 (87.72%) | 497 (1.61%) |
| Time-varying predictors [b] | | | |
|     Centrally inserted central catheter (CICC) placed since last LM | Binary | 135562 (56.13%) | 0 (0%) |
|     Tunneled cuffed-centrally inserted central catheter (Tc-CICC) or totally implanted vascular access devices (TIVAD) placed since last LM | Binary | 95315 (39.47%) | 0 (0%) |
|     Peripherally inserted central catheter (PICC) placed since last LM | Binary | 13794 (5.71%) | 0 (0%) |
|     Total lumens of all catheters placed since last LM | Count | 1 (1, 2) | 12503 (40.51%) |
|     Total parenteral nutrition (TPN) ordered in the 7 days before LM | Binary | 2368 (7.67%) | 0 (0%) |
|     Antibacterials for systemic use ordered in 7 days before LM | Binary | 17802 (57.68%) | 0 (0%) |
|     Antineoplastic agents ordered in 7 days before LM | Binary | 6142 (19.90%) | 0 (0%) |
|     Glasgow Coma Scale (GCS) since last LM | Ordinal | 15 (3, 15) | 29985 (97.16%) |
|     Managed by cardiologist at LM | Binary | 1420 (4.63%) | 186 (0.60%) |
|     Managed by traumatologist at LM | Binary | 1491 (4.86%) | 186 (0.60%) |
|     Managed by pediatrician at LM | Binary | 1866 (6.08%) | 186 (0.60%) |
|     Stayed in the Intensive Care Unit at LM | Binary | 5055 (16.38%) | 0 (0%) |
|     Maximum recorded temperature since last LM (°C) | Continuous | 36.70 (36.40, 37.20) | 9333 (30.24%) |
|     Neutropenia observed since last LM | Binary | 494 (2.74%) | 26293 (85.20%) |
|     Urea, last value since previous LM (mg/dL) | Continuous | 3.47 (3.14, 3.85) [c] | 26737 (86.63%) |
|     Creatinine, last value since previous LM (mg/dL) | Continuous | -0.17 ((-0.43, 0.13) [c] | 26709 (86.54%) |
|     Bilirubin, last value since previous LM (mg/dL) | Continuous | -0.76 (-1.20, -0.24) [c] | 28645 (92.82%) |
|     Ciclosporin, last value since previous LM (µg/dL) | Continuous | 4.82 (4.21, 5.32) [c] | 30836 (99.92%) |
|     pH, last value since previous LM | Continuous | 7.39 (7.35, 7.43) [c] | 29876 (96.81%) |

LM: landmark; IQR: interquartile range; N (%): the number and percentage of "yes" for binary variable.
[a] Median (IQR) or N (%) is summarized based on 30,862 catheter episodes for static variables.
[b] Median (IQR) or N (%) is summarized based on 241,501 observations for time-varying variables.
[c] Median (IQR) of the log-transformed values.

Table 2. Overview of imputation methods

| Method | R packages and parameters | Description |
|---|---|---|
| Median/mode imputation | base::median(na.rm = TRUE); base::sort(na.rm = TRUE) | The median for continuous values and mode for categorical values were obtained using non-missing values of each variable in the training set at each landmark timepoint. These medians and modes were then used to impute missing values in the training and validation sets. |
| Last observation carried forward (LOCF) | tidyr::fill(.direction = "down") | For catheter episodes (in training or validation set) with missing values at baseline (landmark 0) imputation was performed using the baseline median/mode using non-missing baseline values in the training set. Then, missing values at subsequent landmarks were imputed with the last known value from the same catheter episode in both the training and validation sets for each variable. |
| Missing indicator method | dplyr::mutate(if_else(is.na(…), 1, 0)) | Missing indicators which are implemented as dummy variables, taking the values of 1 and 0, are added to the multivariable prediction model to indicate whether the values for the corresponding variables are missing [12]. All medical specialty categories share one missing indicator. An arbitrary fixed value of 99 replaces the missing values. |
| Regression imputation (RI) | mice::mice(m = 1, maxit = 1) | Missing values were imputed using regression imputation models fitted on the training set, with the landmark number included as an additional variable in the imputation models. These models were then applied to the validation set to produce a single imputed dataset for both training and validation, without accounting for the uncertainty in the imputed values. |
| Multiple imputation by chained equations without outcome (MICE-xx) | mice::mice(m = 10, maxit = 10) | Multiple imputations were applied using the MICE algorithm to generate multiple completed datasets by iteratively modeling each variable with missing values as a function of the others, including the landmarks [13]. The imputation process was performed on the training set to generate ten imputed datasets, and the resulting imputation models were then used to impute missing values in the validation set, which yielded ten corresponding imputed validation datasets. In MICE-xx, the outcome was excluded from the imputation model during training. Prediction models were trained on each imputed training dataset and applied to the corresponding imputed validation dataset. The final risk predictions were pooled using Rubin's rules to account for imputation uncertainty |
| MICE with outcome (MICE-yx) | mice::mice(m = 10, maxit = 10) | In contrast with MICE-xx, MICE-yx included the outcome as a predictor in the imputation models fitted in the training set. For each patient in the validation set, the outcome was set to missing (consistent with real time prediction) and imputed alongside other missing predictors. Therefore, we also trained an imputation model for the outcome in the training set (even though the outcome was fully observed in the training data). For a patient in the validation data, the imputed outcome values were used to impute the missing values for other predictors. This process mimics a real time deployment |
| | | scenario where the outcome would not be known at the time of prediction. After imputation, the true outcome values were restored for model evaluation. |
| Mixed model approach | lme4::glmer(family = binomial(link = "logit")); lme4::glmer(family = poisson(link = "log")) | Inspired by the 2-stage dynamic landmark model [14], mixed-effects models were fitted using past landmark measurements, which require that all catheter episodes have at least one measurement for each time-varying predictor. Hence, for predictors missing at baseline, we impute their baseline values using the median or mode of the available baseline data. Static predictors were imputed with their median or mode values per landmark. Subsequently, the mixed-effects models were fitted in the training set to capture both the temporal correlation of repeated measurements within catheter episodes and the variation in predictor patterns across catheter episodes. The predicted values from the mixed-effect models replaced the missing values in the training and validation sets. |
| missForestPredict | missForestPredict::missForest(initialization = "median/mode", ntrees = 500, maxiter = 100) | Imputation is done iteratively using random forest imputation models using missForestPredict [15]. Imputation is done using all landmarks and using the landmark number as a variable in the imputation models. The validation set is imputed using the saved imputation models for each variable and for the same number of iterations. |

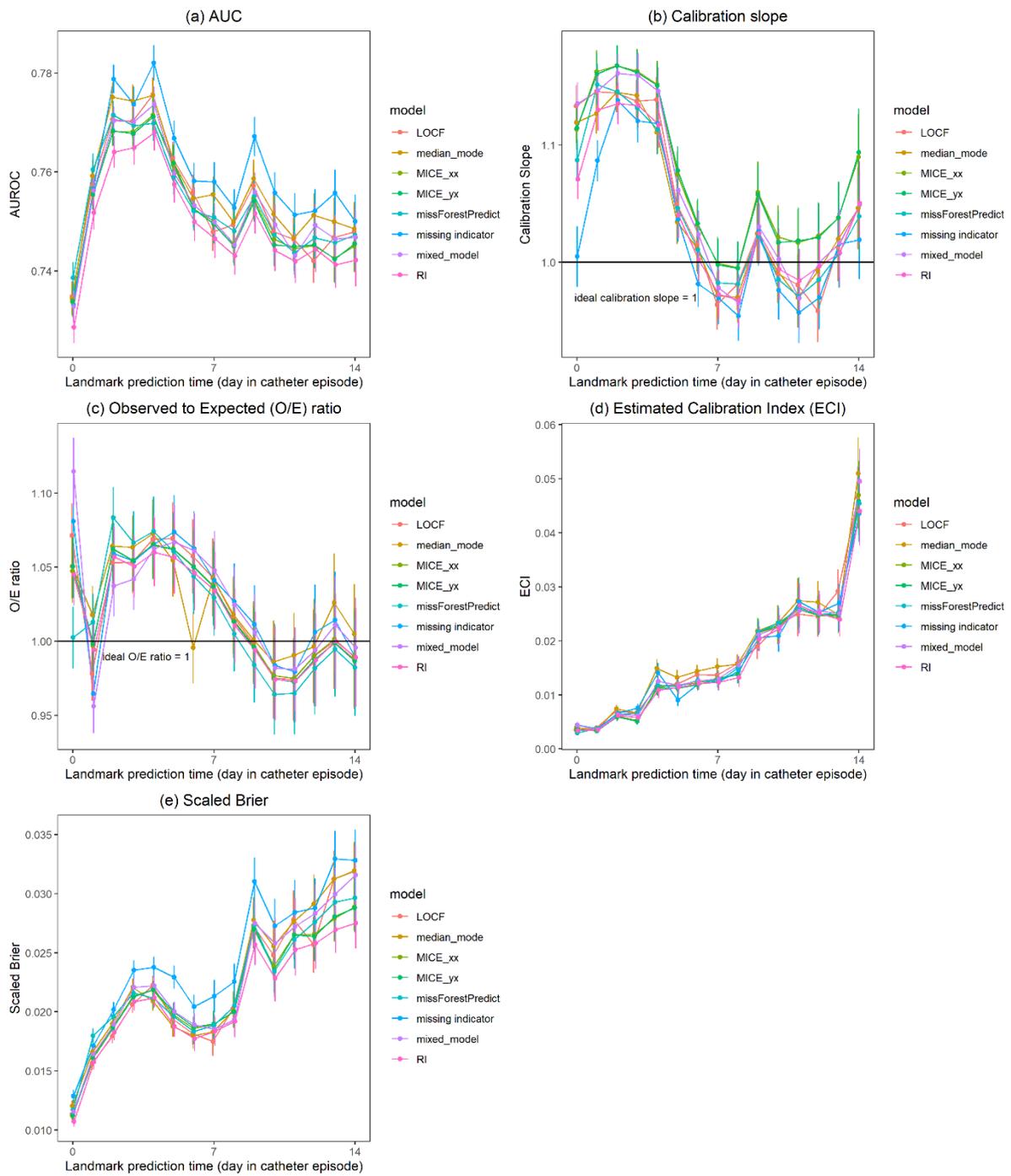

Figure 1 Performance metrics comparison for different imputation methods

Figure 2 Comparison of performance metrics across imputation methods with missing indicators

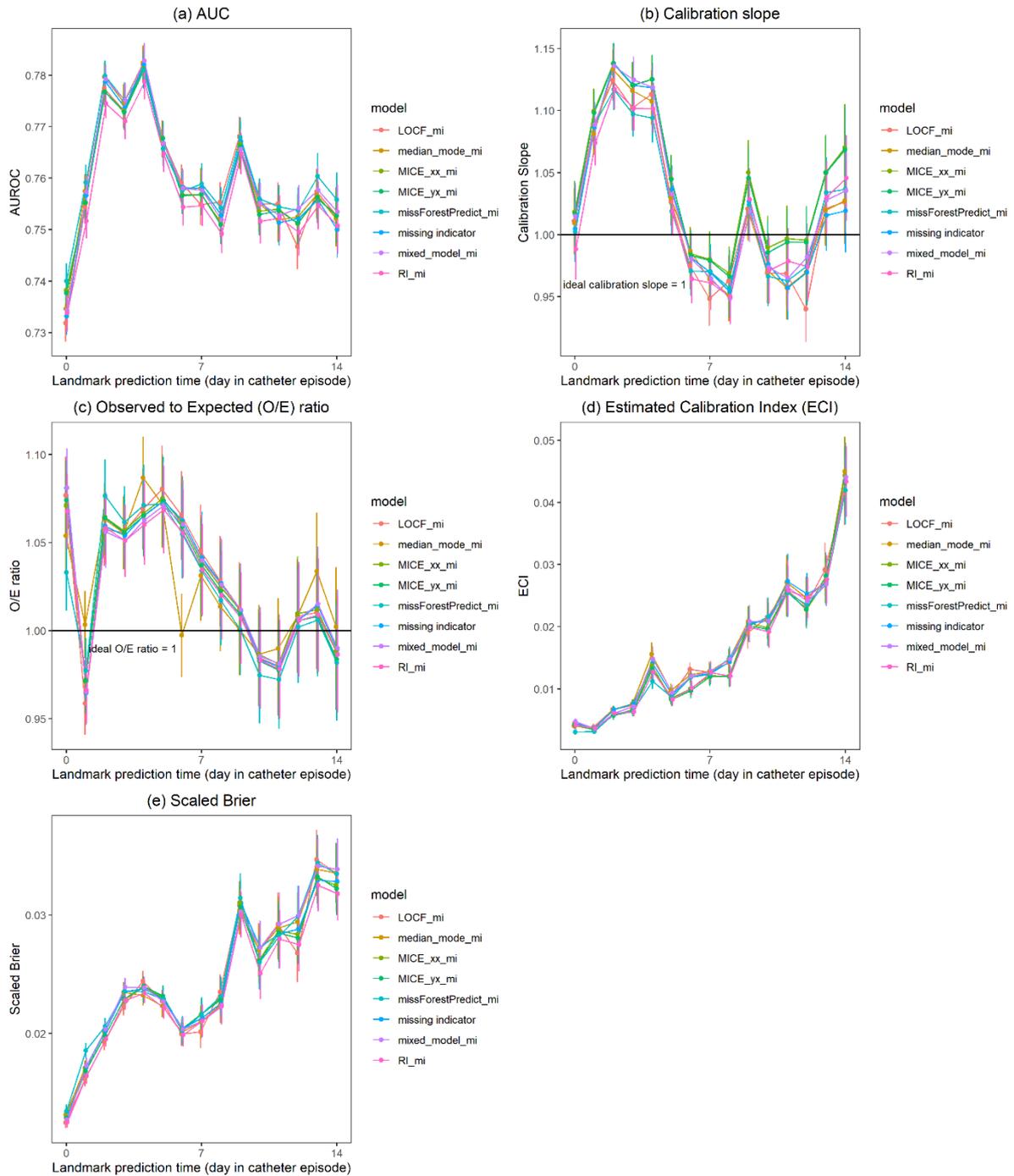

# Supplementary File 1: CLABSI in UZ LEUVEN

The retrospective cohort study consists of patients from the University Hospitals Leuven who were admitted to the hospital and received a catheter between January 2012 and December 2013.

- **Inclusion criteria**

Only hospital stays in participating hospitals with admission starting from January 1st 2012 up to December 30th 2013 were analyzed. Only patients who had a catheter of the following types were included, in accordance with the definitions of CLABSI of the Hospital Hygiene Department:

- Deep venous catheter
- Peripherally inserted central catheter (PICC) (open and valve)
- Midline catheter (open and valve)
- Tunneled dialysis catheter
- Non-tunneled dialysis catheter
- Port-a-cath
- Hickman catheter

Excluded were: rapid infusion system (RIS) catheters, Swan-Ganz catheters, umbilical catheter, Coolgard catheter introducers, pacemakers, arterial catheter, peripheral venous catheter, Extracorporeal membrane oxygenation (ECMO), intra-aortic balloon pump (IABP) Patient-controlled epidural analgesia (PCEA), Pulse index Contour Continuous Cardiac Output (PICCO) .

Patients that had only a dialysis catheter were included only if they have an ICU admission between January 2012 and December 2013, due to the data extraction constraint.

- **Exclusion criteria**

Patients in the neonatology department were documented using a paper-based workflow before October 2013 and did not have electronic records in the system. Thus hospital admissions for patients under the age of 12 weeks have been excluded from the analysis.

- **Catheter episodes**

Considering that there were situations that patients received more than one catheter simultaneously or consecutively in ICU or in other hospital wards, it is difficult to distinguish the effect of those catheters on the risk of CLABSI. Thus, a scheme was developed to make a difference between these situations.

- When the patient received only one catheter, this was regarded as one observation. Its time at risk is the time interval from the catheter placement to the catheter removal plus 48 hours, according to the definition of CLABSI [1].

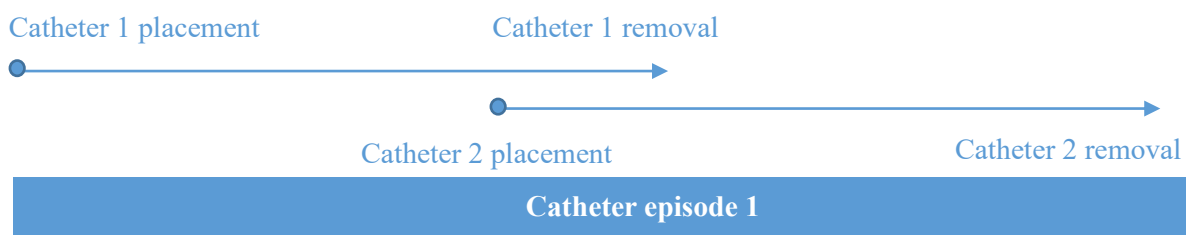

- If the same patient received two catheters and the time interval between these two catheters was less than 48 hours, we treat them as one observation, which means that their time at risk was counted together, from the start of the earlier catheter to the end of the later catheter.

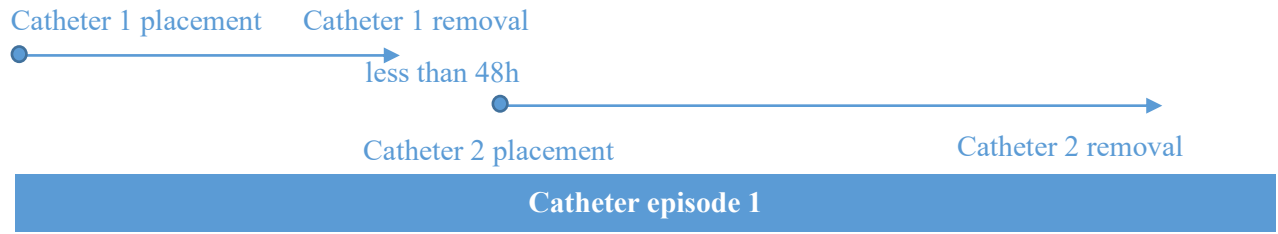

- If the same patients received two catheters and the time interval between these two catheters was more than 48 hours, we treat them as two observations, that is, their time at risk are counted separately.

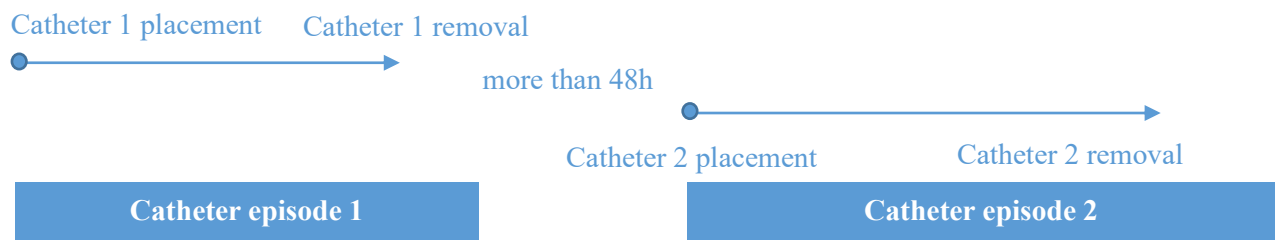

- **Outcome**

There are three types of events (CLABSI, death and discharge) considered in this analysis.

- CLABSI: any laboratory-confirmed bloodstream infection (LCBSI) for a patient with central line or within 48 hours after the central line removal. The CLABSI definition follows the Sciensano definition published in 2019 [2], with specific criteria excluding infection present on admission, secondary infections, skin contamination, and mucosal barrier injury LCBSI. Symptoms criteria are not checked, and it is assumed that cultures are ordered based on relevant symptoms. The time window for secondary infections is not clearly defined in Sciensano, and we use the time window of the 17 days, considering the CLABSI episode length of 14 days plus the time window of 3 days.
- Discharge: hospital discharge or 48 hours after catheter removal, whichever happens first. A patient is still considered at risk within 48 hours after catheter removal.
- Death: Either the first contact with palliative care during admission, transfer to palliative care or patient death, whichever happens first. Patients are not closely monitored in palliative care and predictions for this ward are deemed non-actionable, thus of limited value since there are minimal opportunities to prevent CLABSI events in this context.

# Supplementary File 2: Variable selection

The variable selection follows the procedure:

1. We selected influential predictors without missing values based on domain knowledge and previous work [3]. These predictors were, PAT_age, MED_7d_TPN, MS_is_ICU_unit, MED_L2_7d_L01_ANTINEOPLASTIC_AGENTS, MED_L2_7d_J01_ANTIBACTERIALS_FOR_SYSTEMIC_USE, CAT_catheter_type_binary_all_CVC, CAT_catheter_type_binary_all_Tunneled_CVC, CAT_catheter_type_binary_all_Port_a_cath, CAT_catheter_type_binary_all_PICC.

2. Then we selected influential predictors with missing values based on the Least Absolute Shrinkage and Selection Operator (LASSO) regression results, including laboratory tests and vital signs, scores, number of lumens of various catheter types, admission source and medical specialties. We conducted variable selection based on 100 repeated train-test splits on the data from 2012 to 2013. For each split, we used a random sample of two-thirds of hospital admissions from the landmark dataset, with the remaining one third of the admissions for model validation. Each hospital admission's landmarks and catheter episodes were entirely included in the training set or the test set. Only the training sets are utilized in the LASSO regression. We performed cross-validation (split by hospital admission identifier) to identify the optimal lambda for the LASSO model, subsequently fitting the model with this lambda. Candidate predictors included all predictors with missing values as well as all catheter type predictors in the dataset. We extracted the coefficients from the fitted LASSO model and computed the standard deviation of each predictor, then obtained the standardized coefficients by multiplying the coefficients by the corresponding standard deviations of the predictor variables. Then we assessed the relative importance of the predictors and selected the top 20 important variables. The summarization of the frequency of variables being selected across the 100 runs is shown below:

Table S1 Summarization of frequency of variables selection by LASSO across 100 runs

| variable | Nr |
| --- | --- |
| CARE_NEU_GCS_score_last | 100 |
| CAT_lumens_CVC | 100 |
| LAB_is_neutropenia | 100 |
| LAB_urea_last | 100 |
| MS_medical_specialty_bin_drop_Cardiac | 100 |
| MS_medical_specialty_bin_drop_Traumatology | 100 |
| CAT_lumens_Tunneled_CVC | 92 |
| CAT_catheter_type_binary_all_CVC | 89 |
| MS_medical_specialty_bin_drop_Pediatrics | 87 |
| LAB_bilirubin_last | 86 |
| LAB_creatinine_last | 85 |
| MS_medical_specialty_bin_drop_Abdomen | 74 |
| MS_medical_specialty_bin_drop_Internal_Medicine | 69 |
| CARE_VS_temperature_max | 65 |
| MS_medical_specialty_bin_drop_Neuro | 63 |
| MS_medical_specialty_bin_drop_Hematology | 58 |
| MS_medical_specialty_bin_drop_Endocrinology | 47 |
| MS_medical_specialty_bin_drop_Thoracic_Surgery | 47 |
| MS_medical_specialty_bin_drop_Nephrology | 44 |

| LAB_ciclosporin_last | 43 |
|---|---|
| CAT_catheter_type_binary_all_Tunneled_CVC | 37 |
| ADM_admission_source_binary_all_Home | 35 |
| LAB_pH_last | 34 |

3. Combine CAT_catheter_type_binary_all_Port_a_cath & CAT_catheter_type_binary_all_Tunneled_CVC together as they are both long term catheters

4. Combine CAT_lumens_CVC, CAT_lumens_Tunneled_CVC, CAT_lumens_Dialysis_CVC, CAT_lumens_Port_a_cath, CAT_lumens_PICC together to obtain the total number of lumens

5. According to the sample size calculation in Supplementary File 3.5, at maximum 21 predictors are allowed, assuming a c-statistic of 0.75 and a prevalence of 1.31% for the 7-day CLABSI with two-thirds of admissions from the data. Thus, we further selected the top 3 levels from the predictor MS_medical_specialty due to the limitation of sample size. The final list of predictors used in the model training and validation are:
- "PAT_age"
- "MED_7d_TPN"
- "MS_is_ICU_unit"
- "MED_L2_7d_L01_ANTINEOPLASTIC_AGENTS"
- "MED_L2_7d_J01_ANTIBACTERIALS_FOR_SYSTEMIC_USE"
- "CAT_catheter_type_binary_all_CVC"
- "CAT_catheter_type_binary_all_Tunneled_CVC_Port_a_cath"
- "CAT_catheter_type_binary_all_PICC"
- "CAT_lumens_Total"
- "CARE_NEU_GCS_score_last"
- "CARE_VS_temperature_max"
- "ADM_admission_source_binary_all_Home"
- "LAB_is_neutropenia"
- "LAB_urea_last_log"
- "LAB_creatinine_last_log"
- "LAB_bilirubin_last_log"
- "LAB_ciclosporin_last_log"
- "LAB_pH_last"
- "MS_medical_specialty_bin_drop_Cardiac"
- "MS_medical_specialty_bin_drop_Traumatology"
- "MS_medical_specialty_bin_drop_Pediatrics"

# Supplementary File 3: Details regarding imputation methods

*All imputation methods*

For all imputation methods, the number of lumens per catheter type was imputed separately and then summed to calculate the total number of lumens. A key rule applied during imputation is that since catheter types had no missing values, the lumen count was set to 0 whenever the corresponding catheter type value was 0. For specific catheter types including dialysis CVC and port-a-cath, their lumen counts were predefined based on the dataset: 2 for dialysis CVC and 1 for port-a-cath. After consulting with clinical experts, it was decided to automatically assign these values whenever the respective binary indicator variables (CAT_catheter_type_binary_all_Dialysis_CVC and CAT_catheter_type_binary_all_Port_a_cath) were 1. This was implemented prior to performing data imputation.

*MICE*

Within each train-test split, we generate 10 imputations, leading to 10 completed datasets. Logistic regression was used to impute the variables "LAB_is_neutropenia", "ADM_admission_source_binary_all_Home", "MS_medical_specialty_bin_drop_Cardiac", "MS_medical_specialty_bin_drop_Traumatology", and "MS_medical_specialty_bin_drop_Pediatrics". All other variables were imputed using predictive mean matching (PMM). For the Glasgow Coma Scale (GCS) total score, an ordinal variable with 13 levels, ordinal logistic regression (via the "polyreg" method) was the appropriate choice for imputation. However, due to the significant computational time required (4 hours on a high-performance cluster for m=10), we opted for PMM instead. Numeric laboratory tests, except for "LAB_pH_last", exhibited heavy skewness. To address this, a log transformation was applied prior to imputation. Each MICE imputation was performed after 10 iterations to ensure convergence, as visualized by the convergence plots from one run shown in Figure S1.

Figure S1 Multiple imputation convergence process

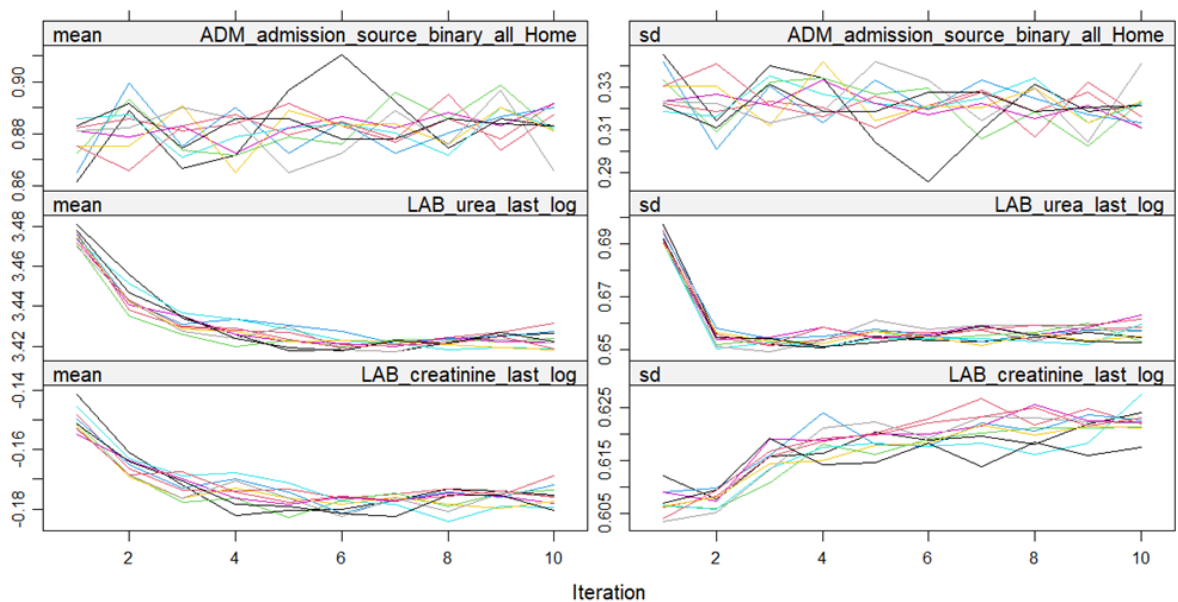

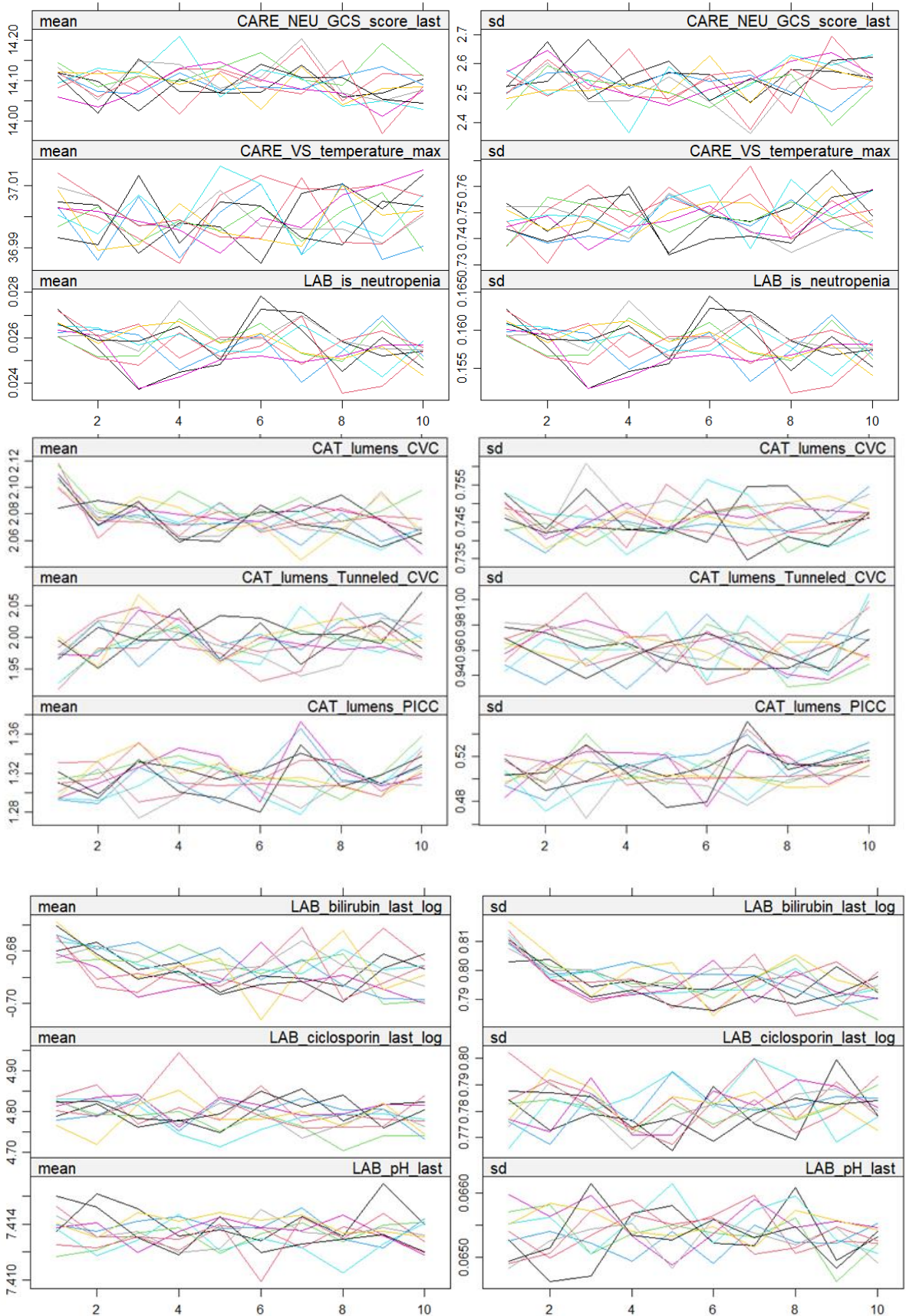

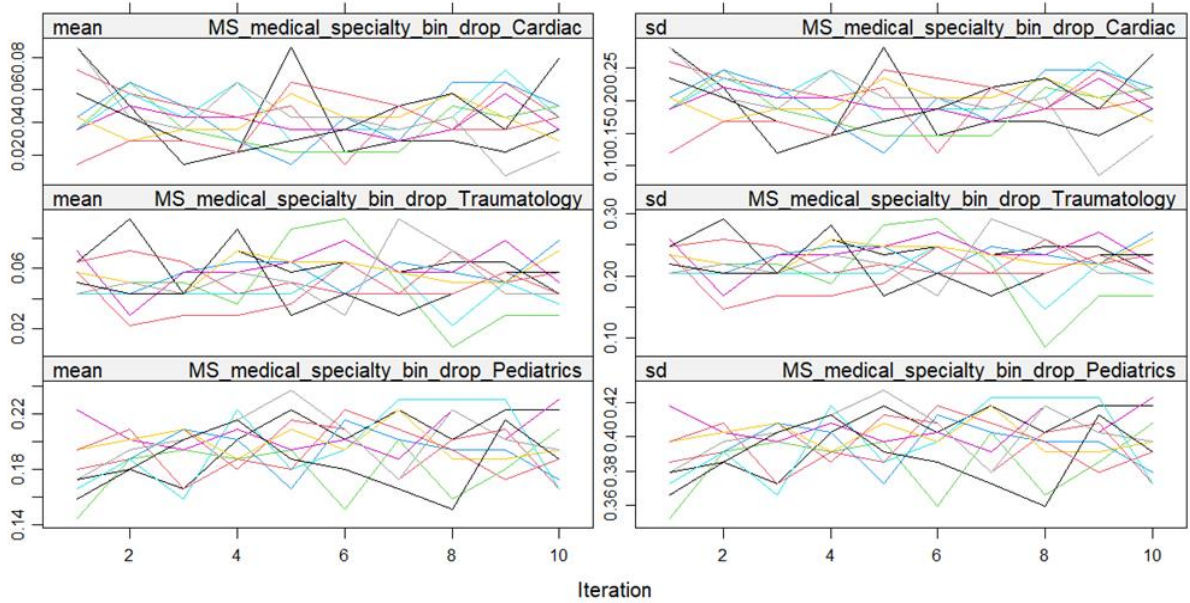

*Mixed model approach*

Similar as MICE, log transformation were applied for all numeric laboratory test variables (except "LAB_pH_last") before fitting the mixed-effect models. The landmark number was also considered as one of the predictors in the mixed-effect model. We assumed a single fixed (population-level) intercept and fixed slopes for each predictor in the mixed-effects models, along with random intercepts for each individual. These random intercepts represent individual-specific deviations from the population average, capturing the correlation among repeated observations within the same individual. Ordinal variables including the GCS total score and the number of lumens per catheter type were fitted with Poisson mixed-effect model, binary variables excluding admission source were fitted with logistic mixed-effect model and continuous variables including the laboratory tests were fitted with linear mixed-effect model. The predicted values replaced the missing values and were used for fitting the model on the training set.

We assumed the separate mixed-effects model for each variable with missing values, except ADM_admission_source_binary_all_Home, which is a baseline variable. It was imputed with mode value due to its low percentage of missingness (1.61%).

Let $i$ represent each catheter episode per individual, and $j$ represent each observation for individual $i$, The model is specified as follows, taking the binary variable LAB_is_neutropenia$_{ij}$ as an example:

logit(Pr(LAB_is_neutropenia$_{ij}$=1)) = $a_1$+$b_1$·LM$_{ij}$ + $c_1$·PAT_age$_i$+$d_1$·MED_7d_TPN$_{ij}$+ $e_1$·MS_is_ICU_unit$_{ij}$+$f_1$·MED_L2_7d_L01_ANTINEOPLASTIC_AGENTS$_{ij}$+$g_1$·MED_L2_7d_J01_ANTIBACTERIALS_FOR_SYSTEMIC_USE$_{ij}$ + $u_{1i}$ where $a_1$ is the fixed intercept. $b_1$, $c_1$, … , $g_1$ are the fixed slopes for each predictor variable. $u_{1i} \sim N(0, \sigma_{u1}^2)$ is the random intercept for individual $i$, accounting for the correlation of repeated measurements within the same individual.

Similar as LAB_is_neutropenia$_{ij}$, the model specification for other variables are:

logit(Pr(MS_medical_specialty_bin_drop_Cardiac$_{ij}$=1)) = $a_2$+$b_2$·LM$_{ij}$ + $c_2$·PAT_age$_i$+$d_2$·MED_7d_TPN$_{ij}$+

$e_2 \cdot MS\_is\_ICU\_unit_{ij} + f_2 \cdot MED\_L2\_7d\_L01\_ANTINEOPLASTIC\_AGENTS_{ij} + g_2 \cdot MED\_L2\_7d\_J01\_ANTIBACTERIALS\_FOR\_SYSTEMIC\_USE_{ij} + u_{2i}$

$logit(Pr(MS\_medical\_specialty\_bin\_drop\_Traumatology_{ij}=1)) = a_3 + b_3 \cdot LM_{ij} + c_3 \cdot PAT\_age_i + d_3 \cdot MED\_7d\_TPN_{ij} + e_3 \cdot MS\_is\_ICU\_unit_{ij} + f_3 \cdot MED\_L2\_7d\_L01\_ANTINEOPLASTIC\_AGENTS_{ij} + g_3 \cdot MED\_L2\_7d\_J01\_ANTIBACTERIALS\_FOR\_SYSTEMIC\_USE_{ij} + u_{3i}$

$logit(Pr(MS\_medical\_specialty\_bin\_drop\_Pediatrics_{ij}=1)) = a_4 + b_4 \cdot LM_{ij} + c_4 \cdot PAT\_age_i + d_4 \cdot MED\_7d\_TPN_{ij} + e_4 \cdot MS\_is\_ICU\_unit_{ij} + f_4 \cdot MED\_L2\_7d\_L01\_ANTINEOPLASTIC\_AGENTS_{ij} + g_4 \cdot MED\_L2\_7d\_J01\_ANTIBACTERIALS\_FOR\_SYSTEMIC\_USE_{ij} + u_{4i}$

$log(E(CARE\_NEU\_GCS\_score\_last_{ij})) = a_5 + b_5 \cdot LM_{ij} + u_{5i}$

$log(E(CAT\_lumens\_CVC_{ij})) = a_6 + b_6 \cdot LM_{ij} + h_6 \cdot CAT\_catheter\_type\_binary\_all\_CVC_{ij} + u_{6i}$

$log(E(CAT\_lumens\_Tunneled\_CVC_{ij})) = a_7 + b_7 \cdot LM_{ij} + k_7 \cdot CAT\_catheter\_type\_binary\_all\_Tunneled\_CVC_{ij} + u_{7i}$

$log(E(CAT\_lumens\_PICC_{ij})) = a_8 + b_8 \cdot LM_{ij} + l_8 \cdot CAT\_catheter\_type\_binary\_all\_PICC_{ij} + u_{8i}$

$CARE\_VS\_temperature\_max_{ij} = a_9 + b_9 \cdot LM_{ij} + c_9 \cdot PAT\_age_i + d_9 \cdot MED\_7d\_TPN_{ij} + e_9 \cdot MS\_is\_ICU\_unit_{ij} + f_9 \cdot MED\_L2\_7d\_L01\_ANTINEOPLASTIC\_AGENTS_{ij} + g_9 \cdot MED\_L2\_7d\_J01\_ANTIBACTERIALS\_FOR\_SYSTEMIC\_USE_{ij} + u_{9i} + \varepsilon_{9ij}$

$LAB\_urea\_last_{ij} = a_{10} + b_{10} \cdot LM_{ij} + c_{10} \cdot PAT\_age_i + d_{10} \cdot MED\_7d\_TPN_{ij} + e_{10} \cdot MS\_is\_ICU\_unit_{ij} + f_{10} \cdot MED\_L2\_7d\_L01\_ANTINEOPLASTIC\_AGENTS_{ij} + g_{10} \cdot MED\_L2\_7d\_J01\_ANTIBACTERIALS\_FOR\_SYSTEMIC\_USE_{ij} + u_{10i} + \varepsilon_{10ij}$

$LAB\_creatinine\_last_{ij} = a_{11} + b_{11} \cdot LM_{ij} + c_{11} \cdot PAT\_age_i + d_{11} \cdot MED\_7d\_TPN_{ij} + e_{11} \cdot MS\_is\_ICU\_unit_{ij} + f_{11} \cdot MED\_L2\_7d\_L01\_ANTINEOPLASTIC\_AGENTS_{ij} + g_{11} \cdot MED\_L2\_7d\_J01\_ANTIBACTERIALS\_FOR\_SYSTEMIC\_USE_{ij} + u_{11i} + \varepsilon_{11ij}$

$LAB\_bilirubin\_last_{ij} = a_{12} + b_{12} \cdot LM_{ij} + c_{12} \cdot PAT\_age_i + d_{12} \cdot MED\_7d\_TPN_{ij} + e_{12} \cdot MS\_is\_ICU\_unit_{ij} + f_{12} \cdot MED\_L2\_7d\_L01\_ANTINEOPLASTIC\_AGENTS_{ij} + g_{12} \cdot MED\_L2\_7d\_J01\_ANTIBACTERIALS\_FOR\_SYSTEMIC\_USE_{ij} + u_{12i} + \varepsilon_{12ij}$

$LAB\_ciclosporin\_last_{ij} = a_{13} + b_{13} \cdot LM_{ij} + c_{13} \cdot PAT\_age_i + d_{13} \cdot MED\_7d\_TPN_{ij} + e_{13} \cdot MS\_is\_ICU\_unit_{ij} + f_{13} \cdot MED\_L2\_7d\_L01\_ANTINEOPLASTIC\_AGENTS_{ij} + g_{13} \cdot MED\_L2\_7d\_J01\_ANTIBACTERIALS\_FOR\_SYSTEMIC\_USE_{ij} + u_{13i} + \varepsilon_{13ij}$

$LAB\_pH\_last_{ij} = a_{14} + b_{14} \cdot LM_{ij} + c_{14} \cdot PAT\_age_i + d_{14} \cdot MED\_7d\_TPN_{ij} + e_{14} \cdot MS\_is\_ICU\_unit_{ij} + f_{14} \cdot MED\_L2\_7d\_L01\_ANTINEOPLASTIC\_AGENTS_{ij} + g_{14} \cdot MED\_L2\_7d\_J01\_ANTIBACTERIALS\_FOR\_SYSTEMIC\_USE_{ij} + u_{14i} + \varepsilon_{14ij}$

*missForestPredict*

For the variables with missing values, missing data imputations have been performed on each training set using the missForestPredict, which is an adaptation of the missForest algorithm for prediction settings [4]. The missing values were first imputed with median/mode and then iteratively imputed using random forest models until the algorithm converged or for a maximum of 10 iterations. The outcome was not included in the imputation and test datasets were imputed using the missForestPredict imputation models learned on the matching training dataset to mimic prediction model use at the bedside. The out-of-bag (OOB) normalized mean square errors (NMSE) for the variables in missForest and missForest_mi are presented in Figure S2.

Figure S2 Out-of-bag normalized mean square errors

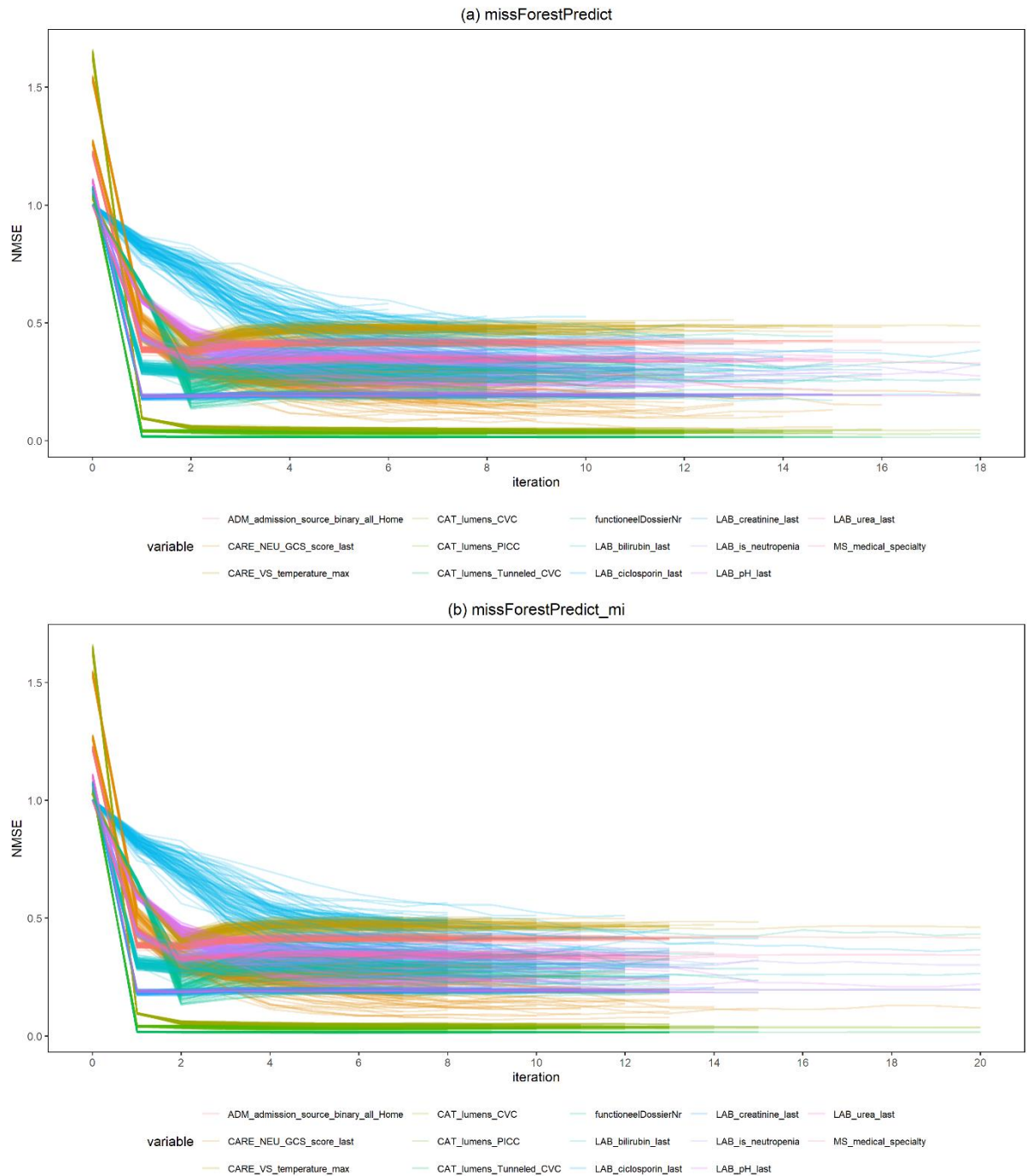

# Supplementary File 4: Distributions of laboratory tests before and after log-transformation

Figure S3 Distribution plots of laboratory tests before and after log-transformation

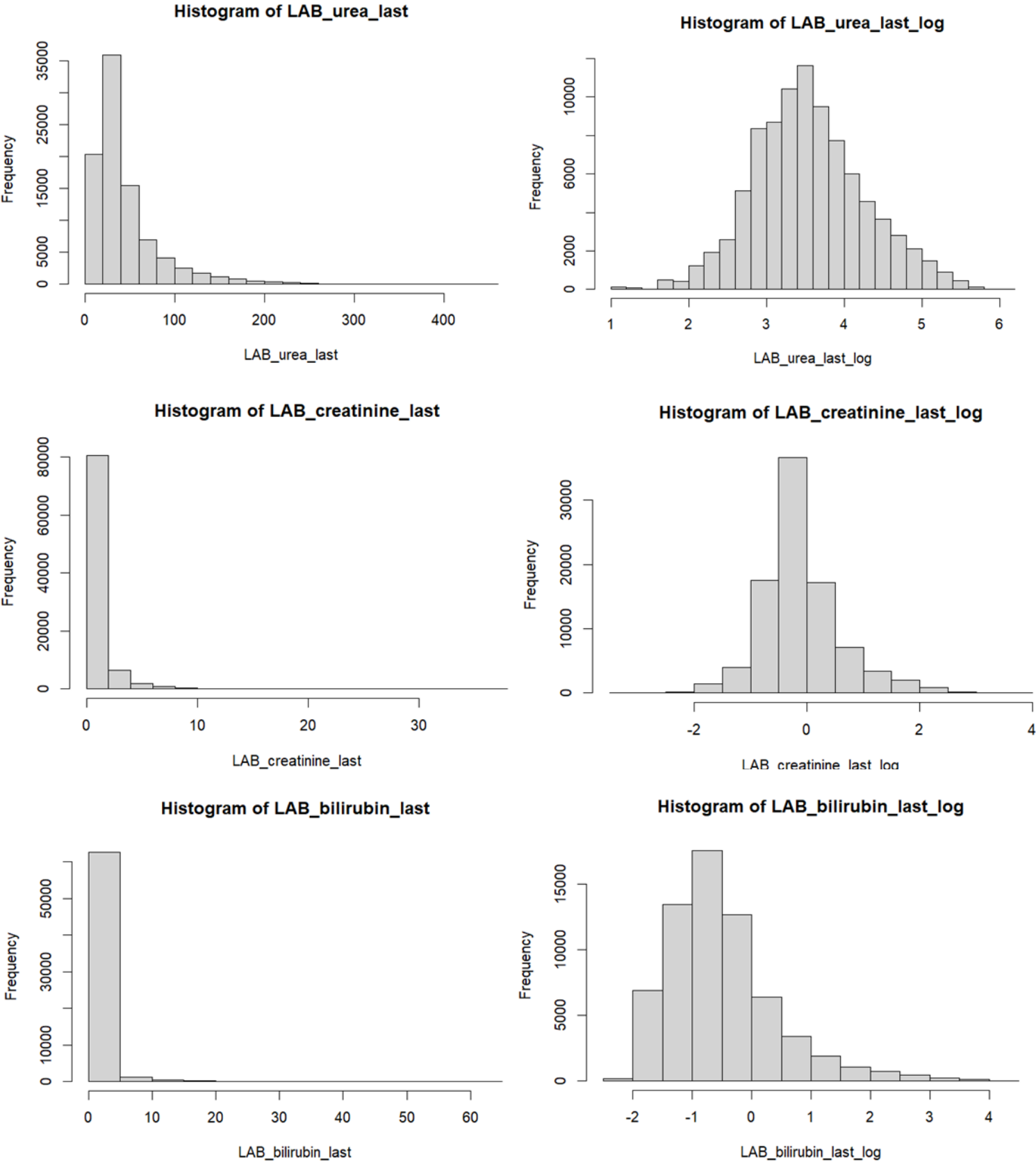

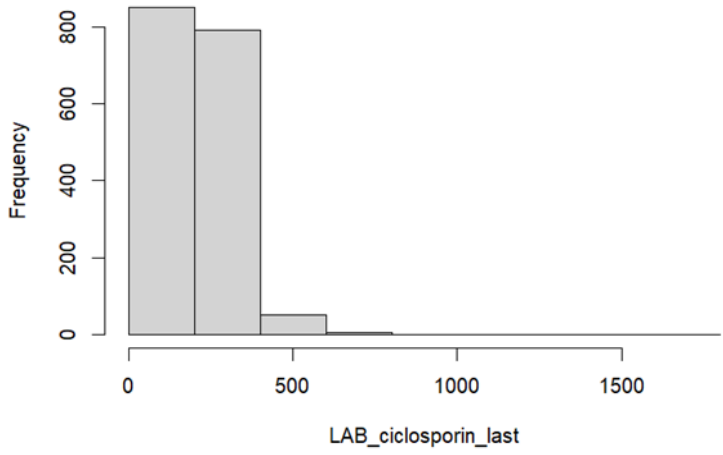
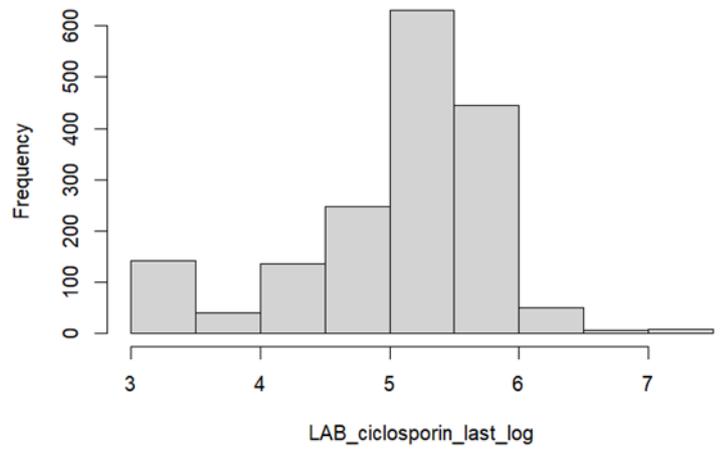
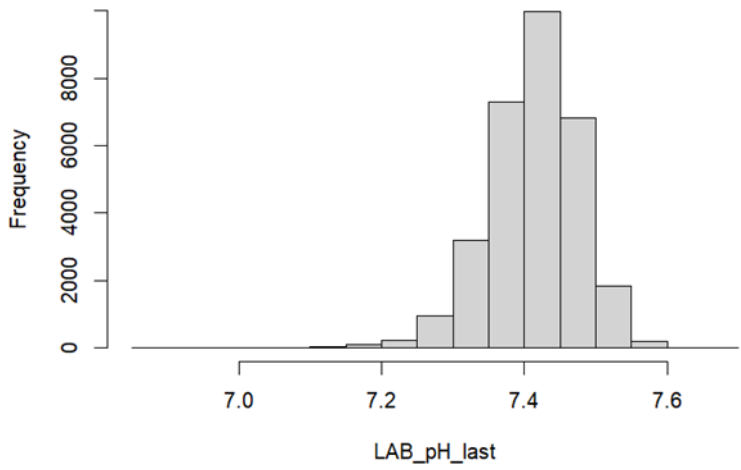

# Supplementary File 5: Landmarking approach

The landmarking approach for dynamic prediction of survival was initially described by van Houwelingen [5]. In brief, at a given landmark time *s* where a prediction is to be made, the data are restricted to individuals who have not yet experienced the event. Predictor values available up to the landmark time are used as covariates in a model for the probability of survival up to some time horizon, conditional on survival to the landmark time. Typically, the focus is on survival to a single time horizon *w*, and censoring is imposed at *w* so that only events up to that time *s+w* are used in the survival analysis. In this approach, the dataset is transformed into multiple censored datasets based on predefined *s* and *w*. Traditionally, a separate cox proportional hazards model can be applied to each landmark dataset and predictions can then be made at each landmark time point. Moreover, a supermodel can be fitted on the stacked super dataset and the landmark supermodel may combine these models by introducing smoothing to permit risk prediction at any landmark. Then dynamic risk prediction can be performed by using the most up-to-date value of patients' covariate values.

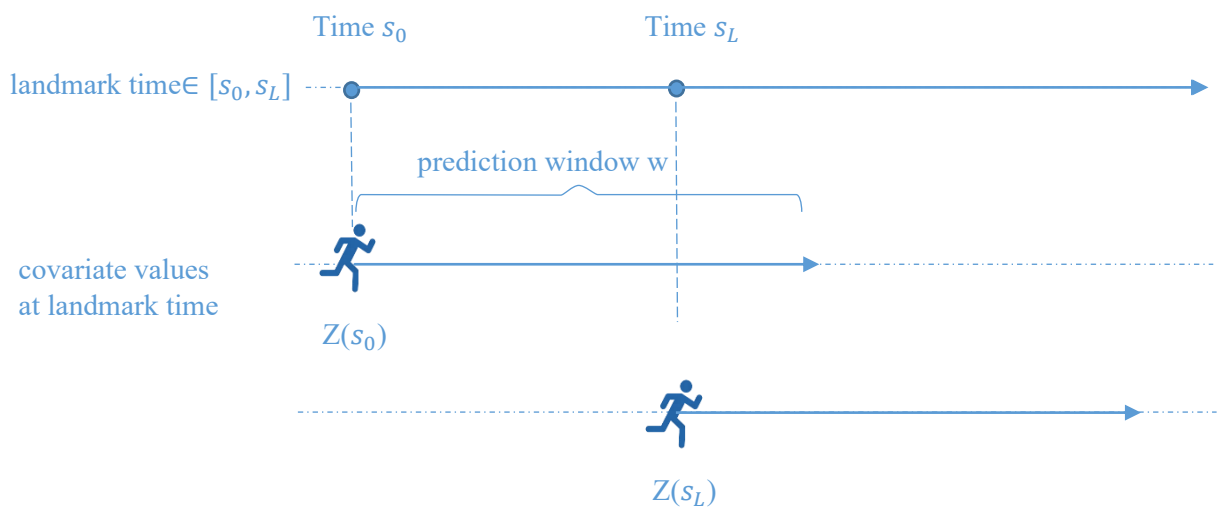

To fit a landmark supermodel [5,6], a stacked dataset is constructed by: (i) firstly selecting a set of landmark points *s* from $[s_0, s_L]$; (ii) then creating a landmark subset by selecting the subjects who have not yet failed from any cause at *s* and adding an administrative censoring at the prediction horizon *s+w*; (iii) finally stacking all the individual landmark subsets into a super prediction dataset.

The following table is an example of the stacked super dataset (with pseudo-anonymization) which is used to develop the dynamic models. The original data is not allowed to share, thus sensitive information are replaced with manually created synthetic data in the following table.

Table S2: example data for landmark cause-specific supermodel

| ID | LM | eventtime | type[a] | MS_is_ICU_unit | LAB_urea_last |
|---|---|---|---|---|---|
| 1 | 0 | 4.42 | 1 | 0 | 25 |
| 1 | 1 | 4.42 | 1 | 0 | NA |
| 1 | 2 | 4.42 | 1 | 0 | 26 |
| 1 | 3 | 4.42 | 1 | 1 | NA |
| 1 | 4 | 4.42 | 1 | 0 | 28 |
| 2 | 0 | 7.00[b] | 0 | 1 | 55 |
| 2 | 1 | 8.00[b] | 0 | 1 | 38 |
| 2 | 2 | 9.00[b] | 0 | 1 | 41 |
| 2 | 3 | 9.34 | 1 | 1 | 51 |
| 2 | 4 | 9.34 | 1 | 1 | 52 |
| 2 | 5 | 9.34 | 1 | 1 | NA |
| 2 | 6 | 9.34 | 1 | 1 | 43 |
| 2 | 7 | 9.34 | 1 | 1 | 39 |
| 2 | 8 | 9.34 | 1 | 1 | NA |
| 2 | 9 | 9.34 | 1 | 1 | 57 |
| 3 | 0 | 1.29 | 2 | 1 | 32 |
| 3 | 1 | 1.29 | 2 | 1 | NA |
| 4 | 0 | 4.56 | 3 | 0 | 59 |
| 4 | 1 | 4.56 | 3 | 0 | 47 |
| 4 | 2 | 4.56 | 3 | 0 | 38 |
| 4 | 3 | 4.56 | 3 | 0 | 45 |
| 4 | 4 | 4.56 | 3 | 0 | 41 |

LM: landmark time; eventtime: time when any type of event happened; type: type of the event; MS_is_ICU_unit: binary indicator, whether the patient now (at the exact second of the current LM) is in ICU; LAB_urea_last: continuous variable, last value of urea since previous LM. Unit: mg/dL; NA: missing value which is not recorded.

[a]type=1 (CLABSI); type=2 (Death); type=3 (Discharge); type=0 (Censored)

[b]Discharge here means catheter removal. As prediction time window is 7-day from each landmark time, individuals who are free of event up to 7 days of follow-up from the corresponding LM are administratively censored.

To fit a landmark cause-specific supermodel, we consider the baseline hazards $\lambda_{j0}(t)$ from each of the cause-specific Cox models for event J (J=1,…, j) depends on s and this can be modelled by as $\lambda_{j0}^{cs}(t|s) = \lambda_{j0}^{cs}(t) \exp\left(\gamma_j(s)\right)$ and $\gamma_j(s) = \gamma_{j1}\left(\frac{s}{30}\right) + \gamma_{j2}\left(\frac{s}{30}\right)^2$. Thus, the cause-specific hazards of supermodel for event J from a landmark time $s \in [s_0, s_L]$ for time $t$ ($s \leq t \leq s+w$) is:

$$\lambda_j^{cs}(t|Z(s),s) = \lambda_{j0}^{cs}(t|s) exp\left(\beta_j(s)Z(s)\right) = \lambda_{j0}^{cs}(t) \exp\left(\gamma_j(s) + \beta_j(s)Z(s)\right) \quad (1)$$

Then $w$-day event free survival at any time point s in the window $[s_0, s_L]$ can be estimated with either the exponential approximation [7]:

$$S(s + w|Z(s), s) = \exp\left(-\int_s^{s+w} \sum_{j=1}^{J} \lambda_j^{cs}(t|Z(s),s)\, dt\right) = \exp\left(-\sum_{s \leq t_i \leq s+w} \sum_{j=1}^{J} \lambda_j^{cs}(t_i|Z(s),s)\right) \quad (2)$$

The cause-specific cumulative incidence for event J at any time point s in the window $[s_0, s_L]$ can be obtained with

$$F_j(s + w|Z(s), s) = \int_s^{s+w} \lambda_j^{cs}(t|Z(s),s) S(t|Z(s),s)\, dt = \sum_{s \leq t_i \leq s+w} \lambda_j^{cs}(t_i|Z(s),s) S(t_i|Z(s),s) \quad (3)$$

where $t_i$ are event times in the training dataset used to fitting the landmark cause-specific supermodel.

# Supplementary File 6: Sample size calculation

We used the pmsampsize package in R to estimate the minimum sample size required for developing a static logistic regression model and a static Cox proportional hazards [8]. According to a systematic review of CLABSI prediction models [9], the mean of optimism-corrected AUC is 0.75 in the studies with similar EHR setting. Assuming we can obtain a c-statistic of 0.75 with 21 parameters and a prevalence of 3.14%, a minimum of 7,027 and 6,974 unique catheter episodes are required to develop logistic and Cox models, respectively. When the prevalence is 1.31%, at least 16,621 (logistic) and 15,912 (Cox) catheter episodes are required. Hence, our sample size of 30,862 is sufficiently large for static models if applying a 2:1 train test split. Given the utilization of the landmark approach for dynamic models, where information from adjacent landmarks is leveraged, we conclude that the same sample size rationale is applicable for dynamic models as well.

# Supplementary File 7: Calibration plots

Figure S4 Calibration plots of the landmark cause-specific model for 100 train-test splits per landmark per imputation method

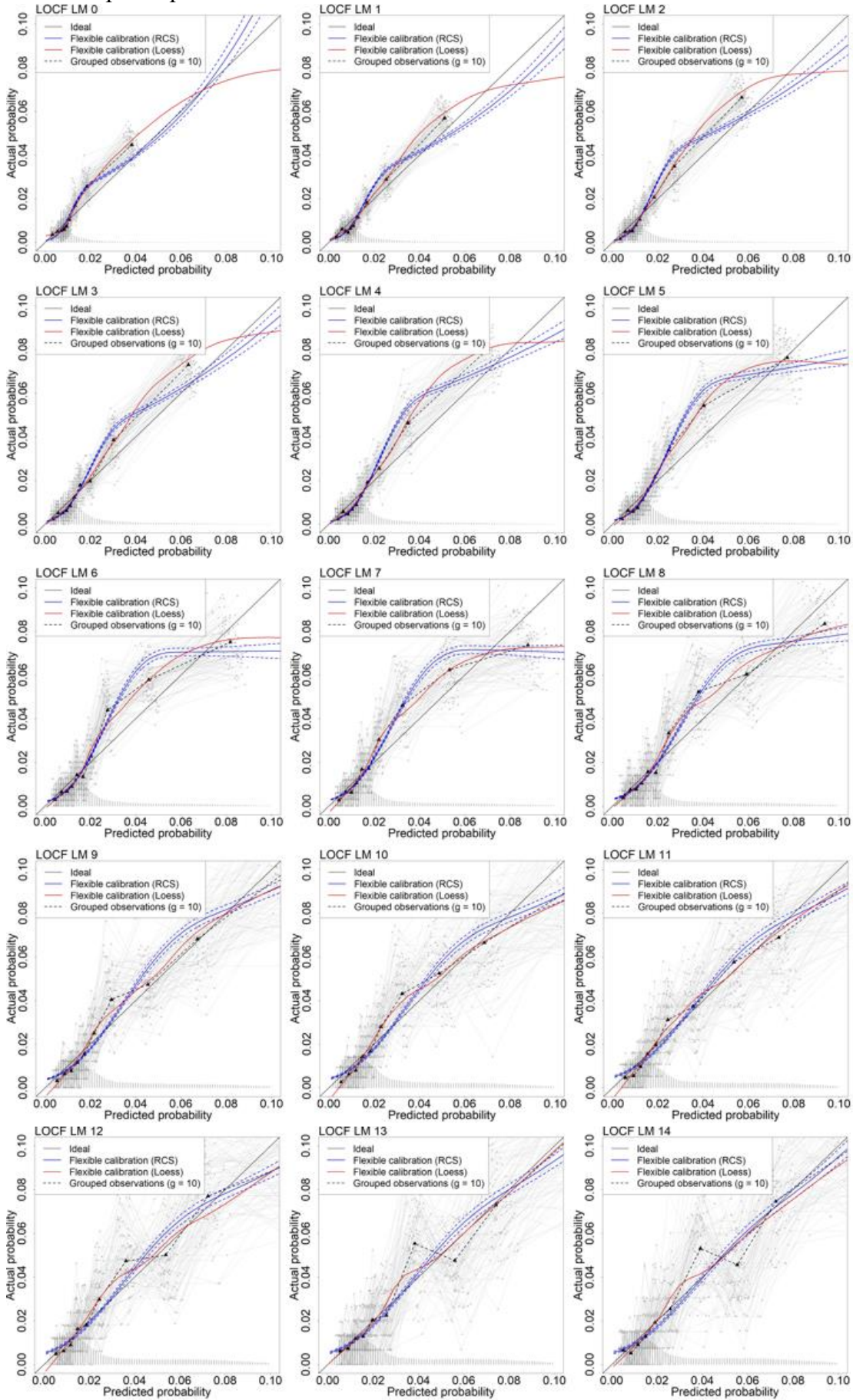

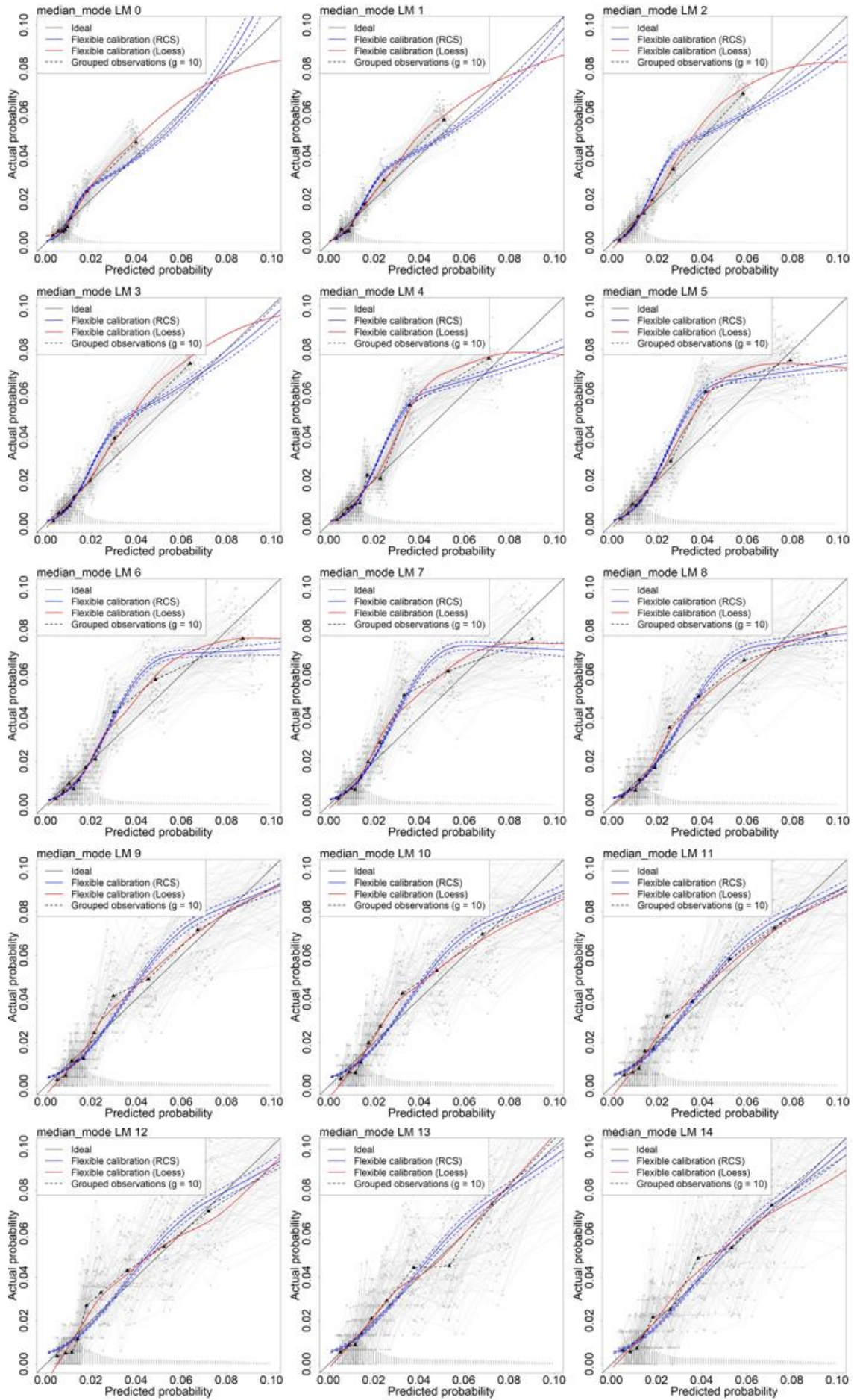

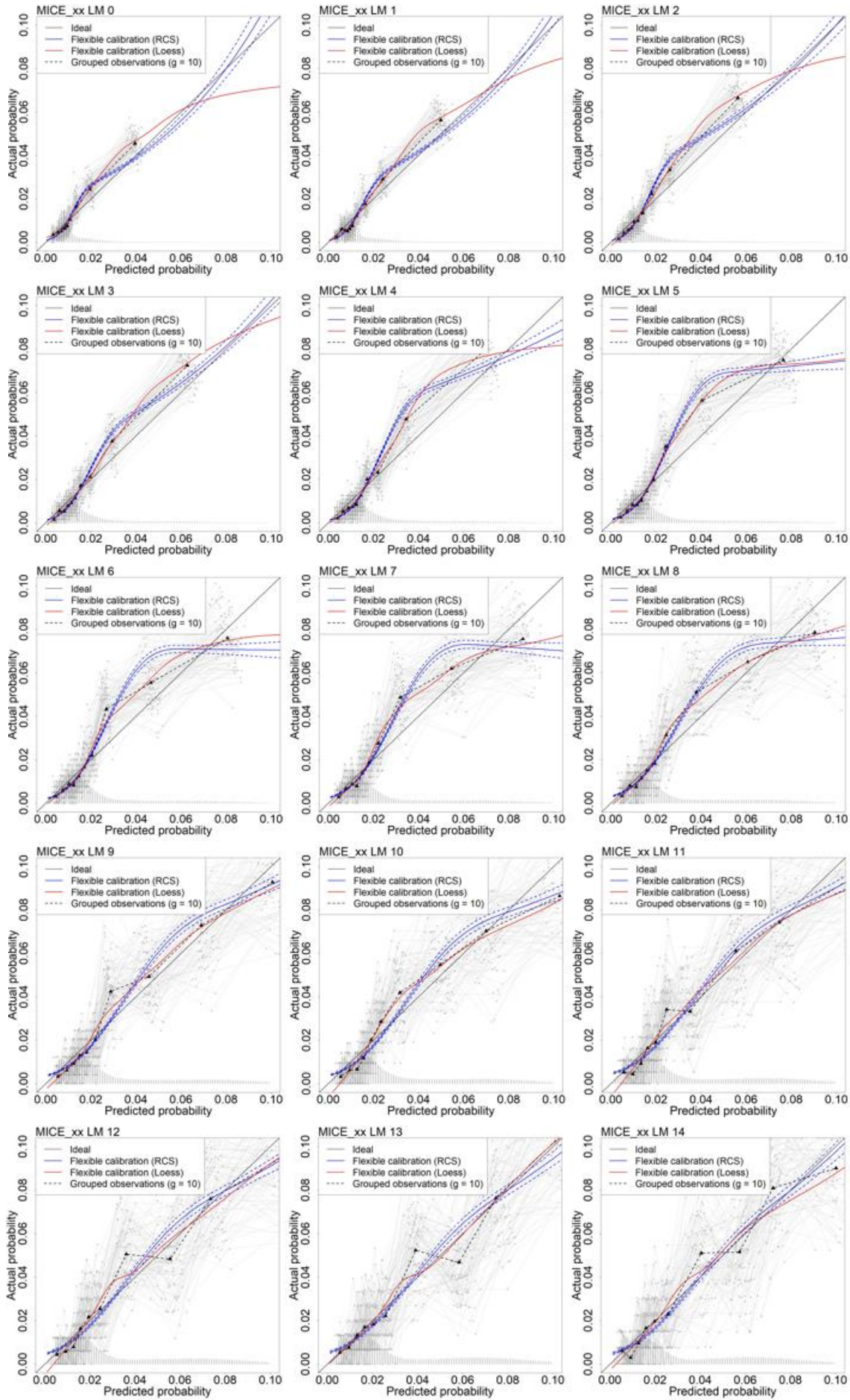

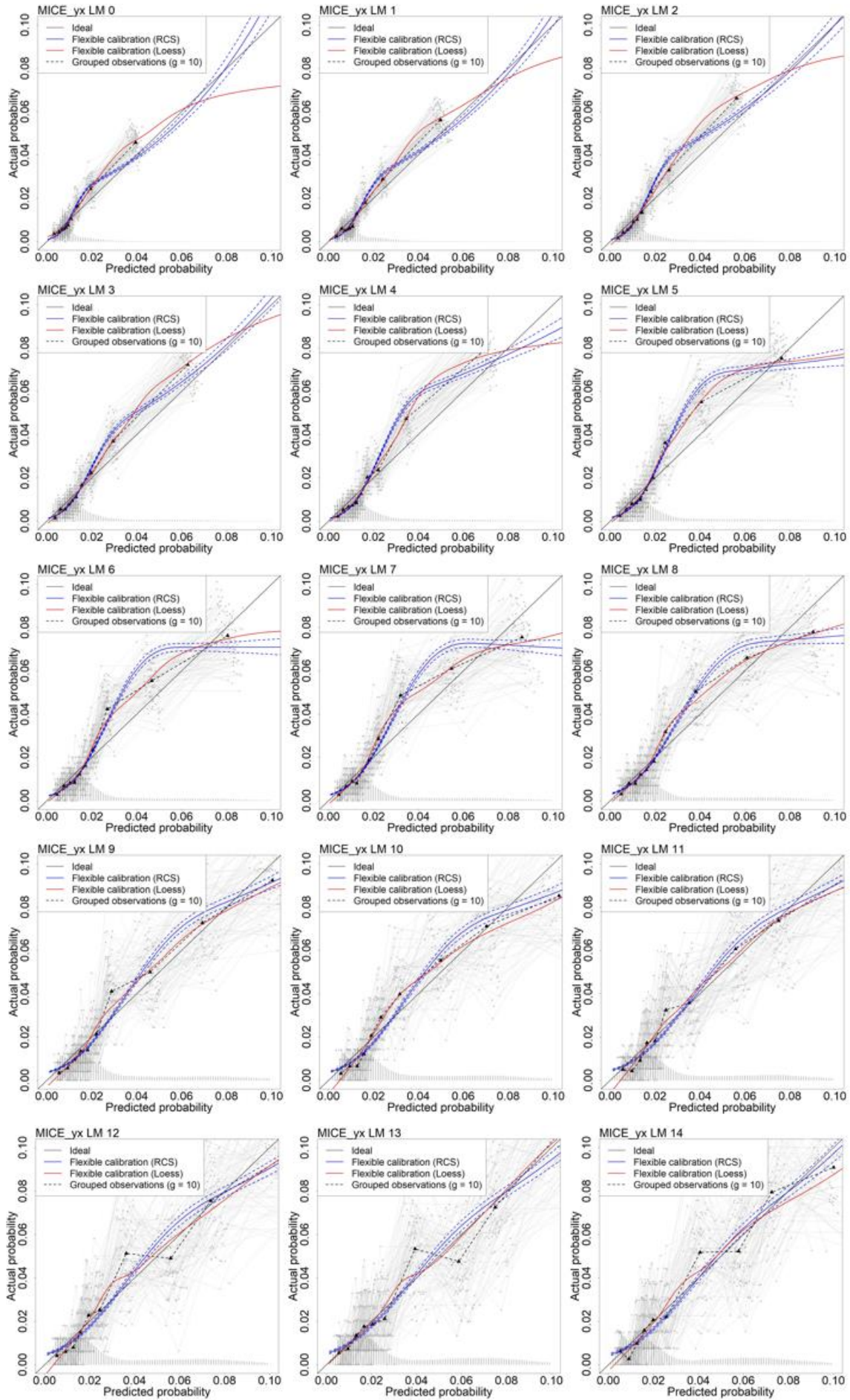

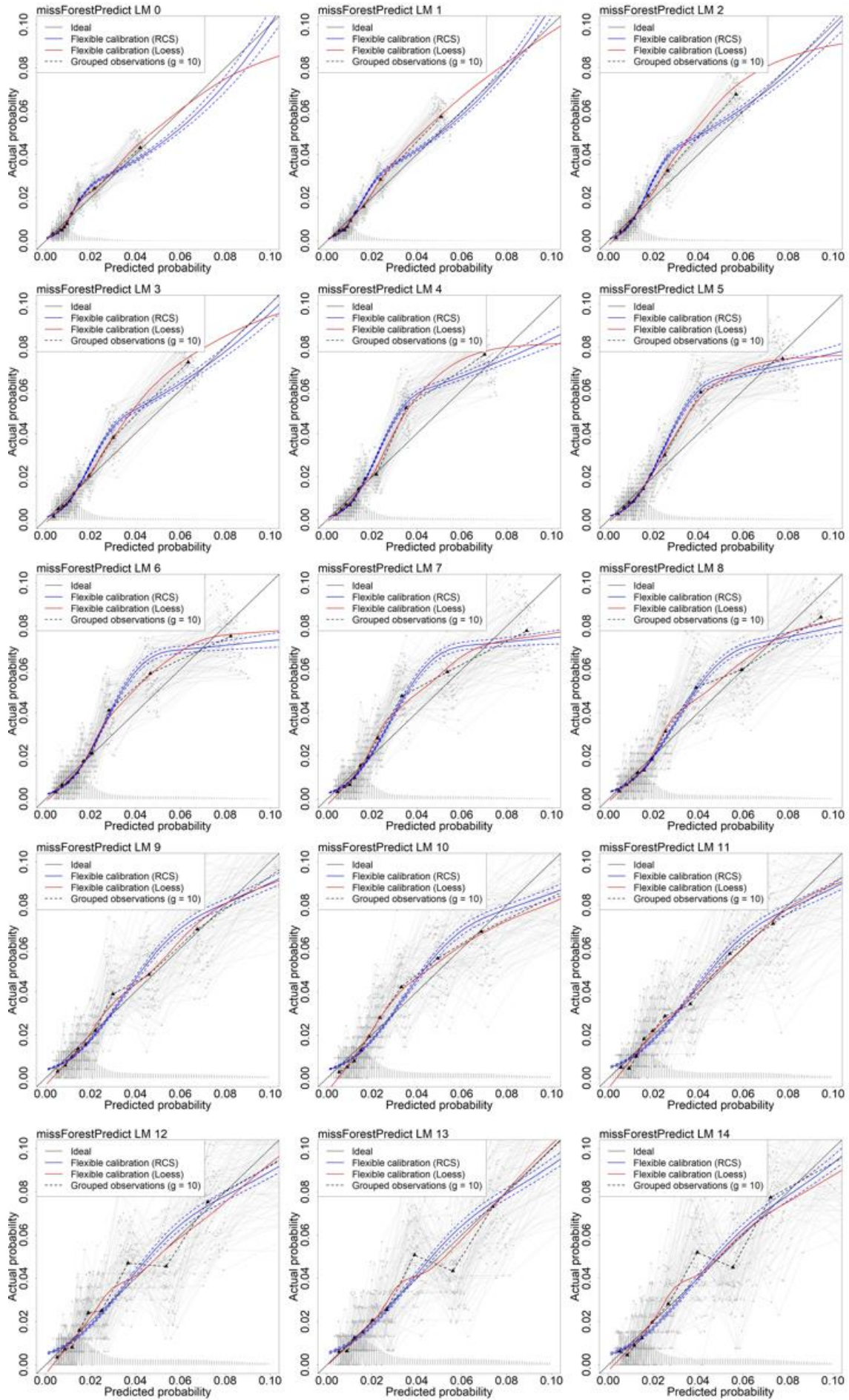

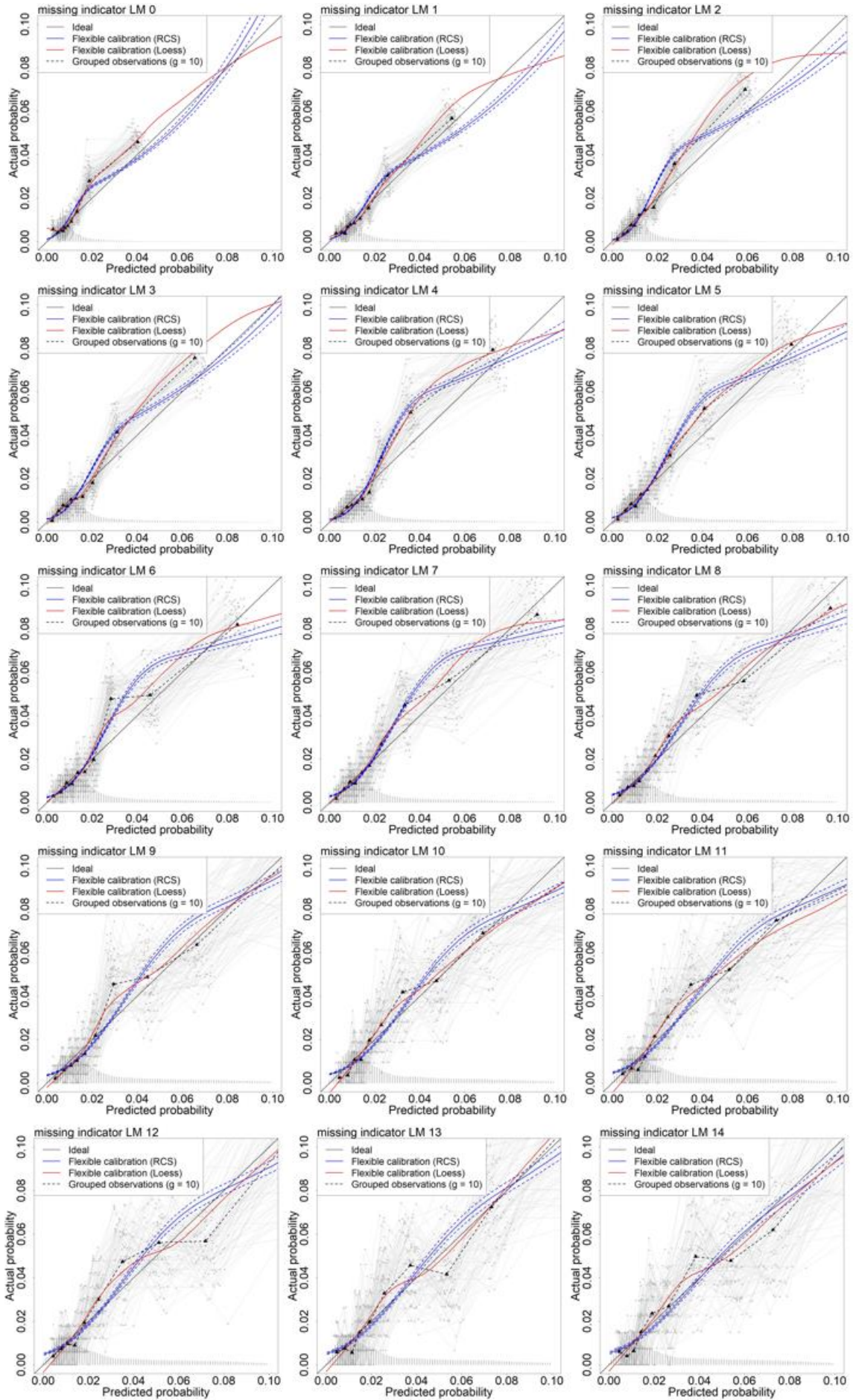

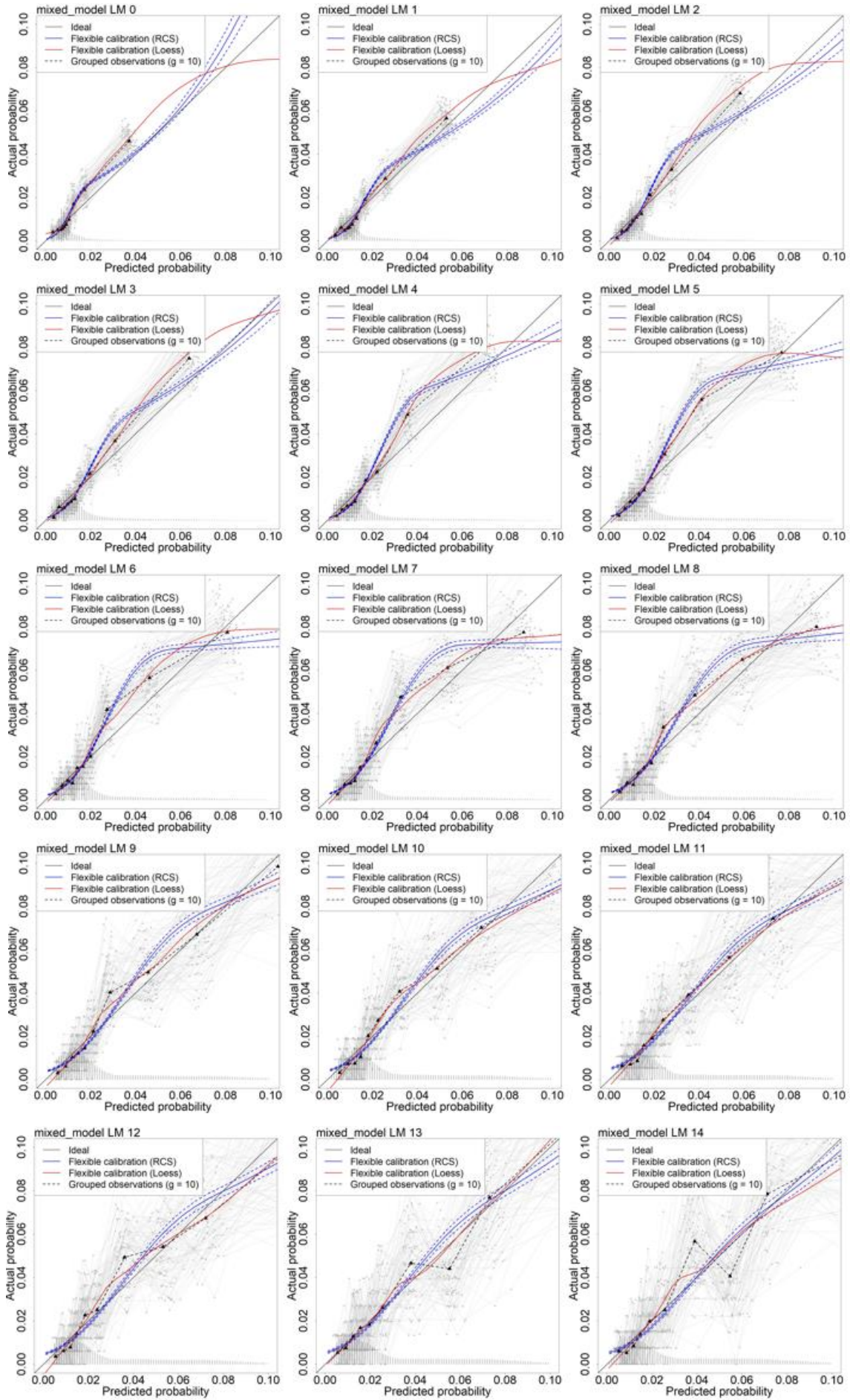

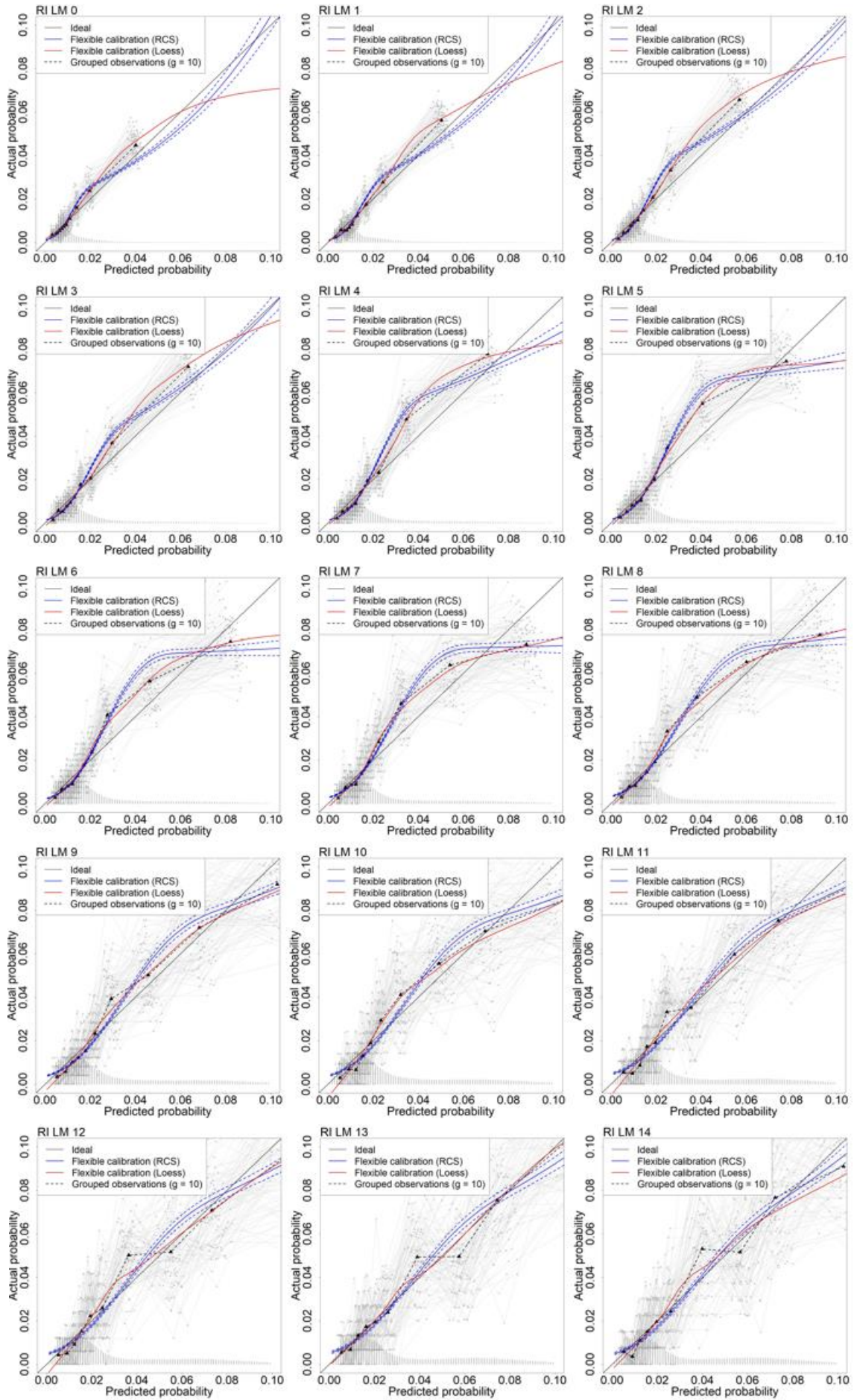

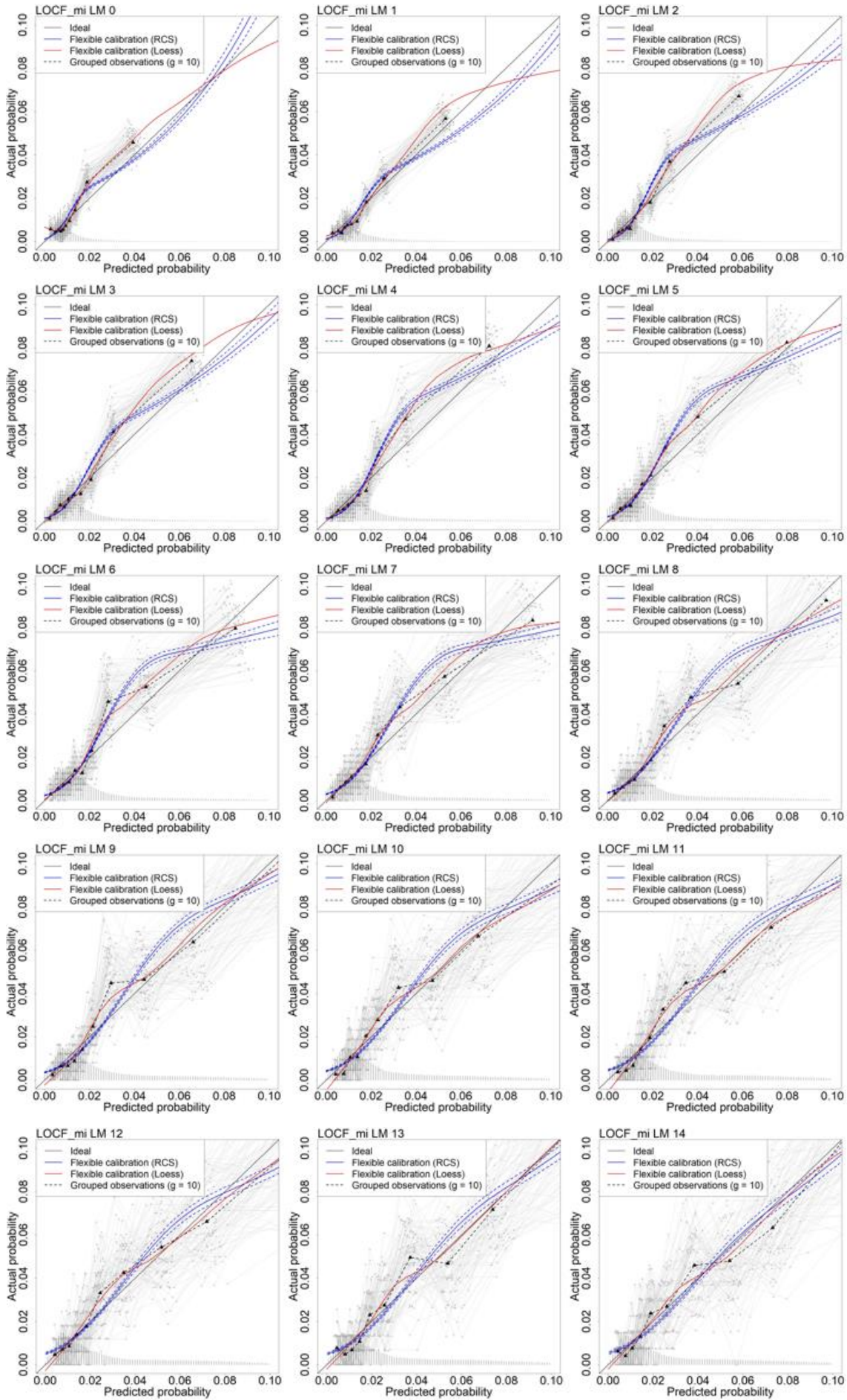

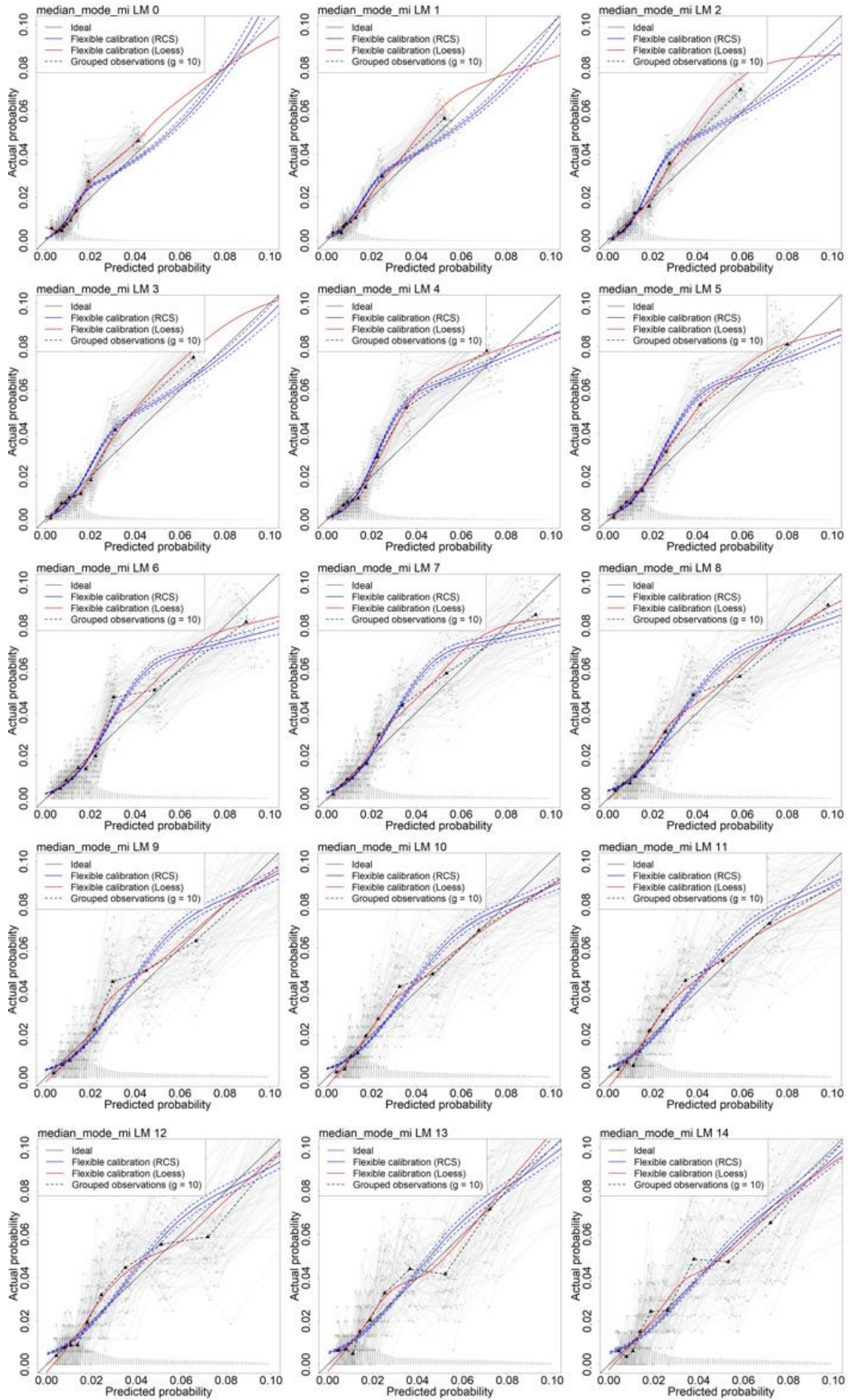

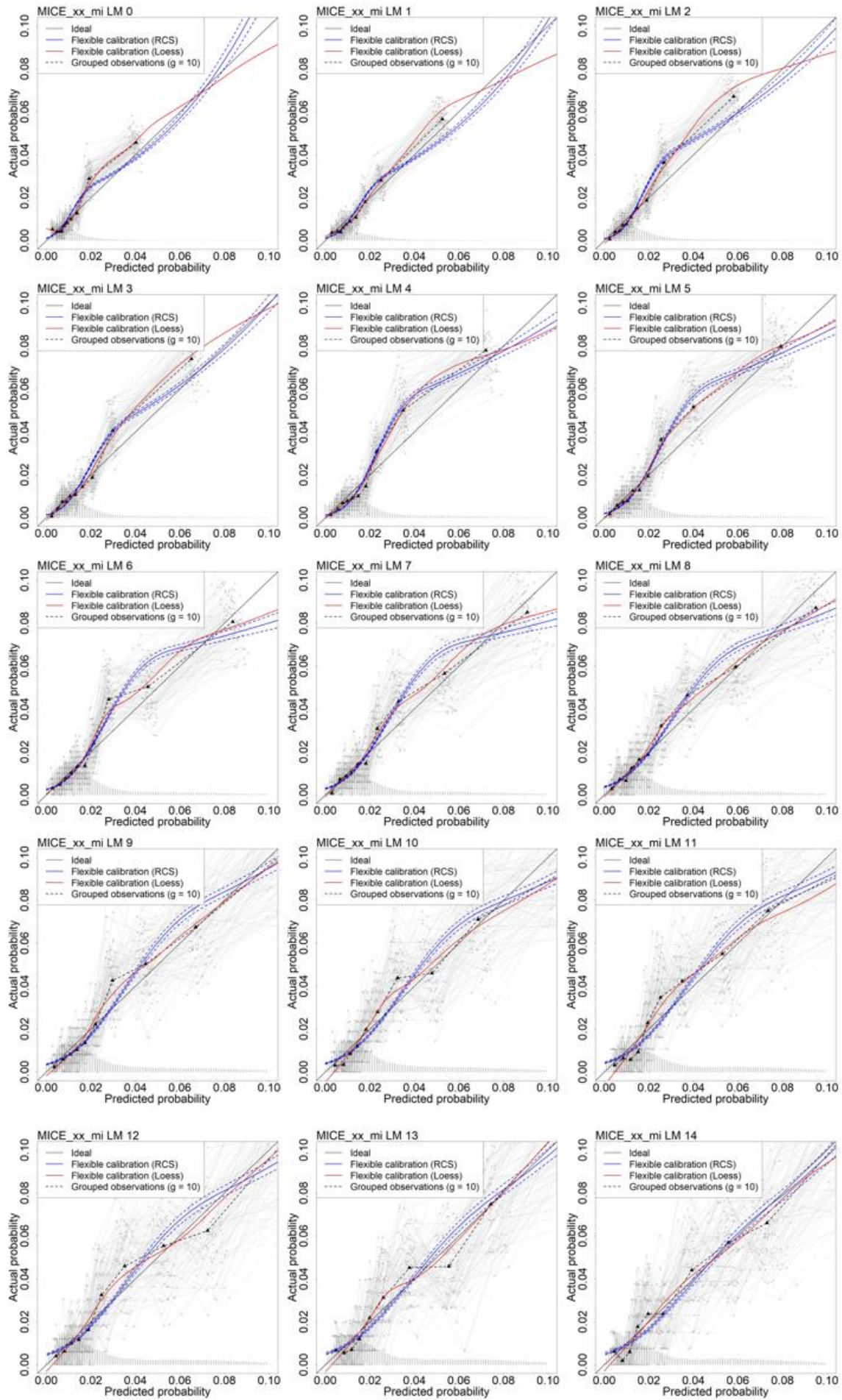

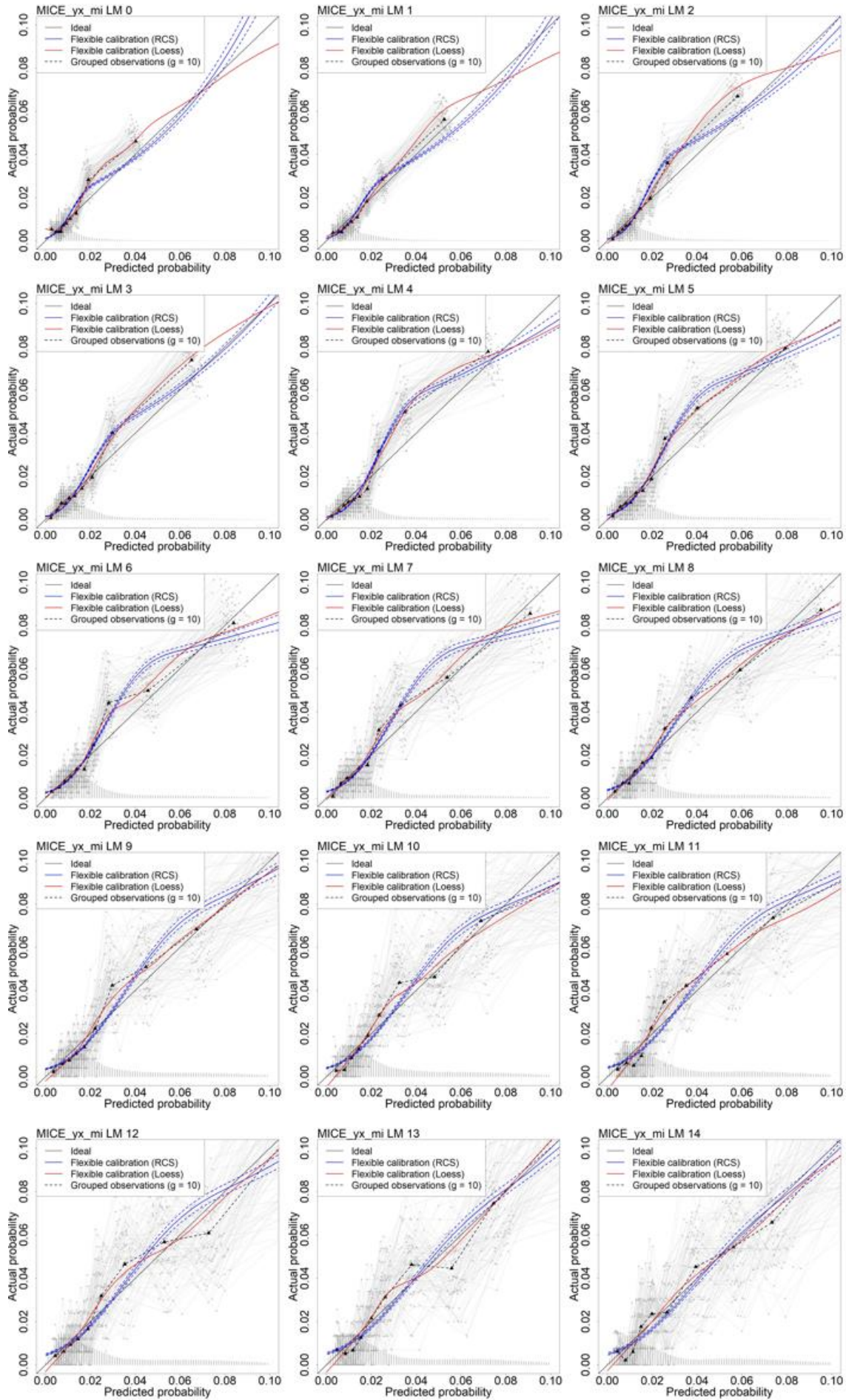

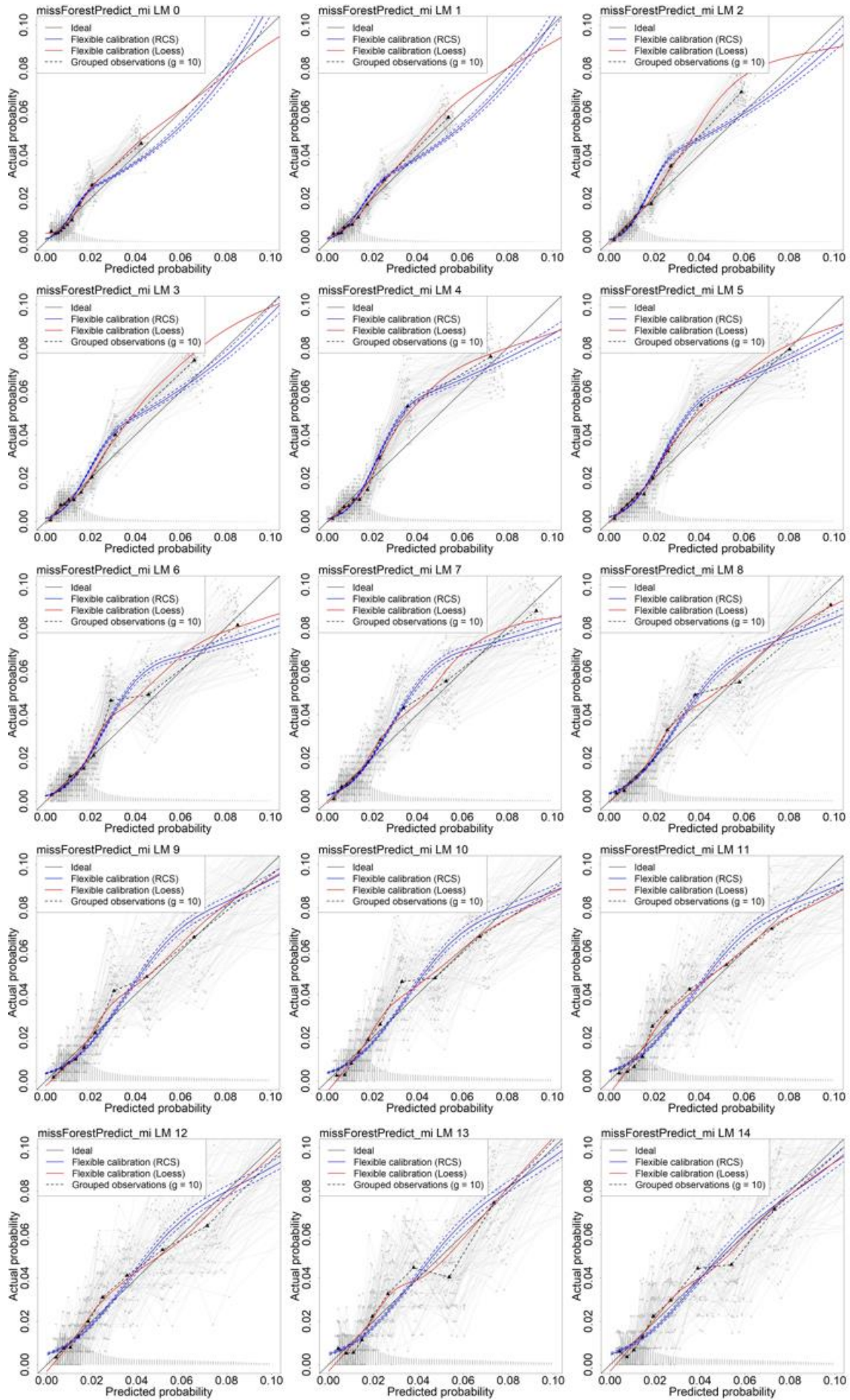

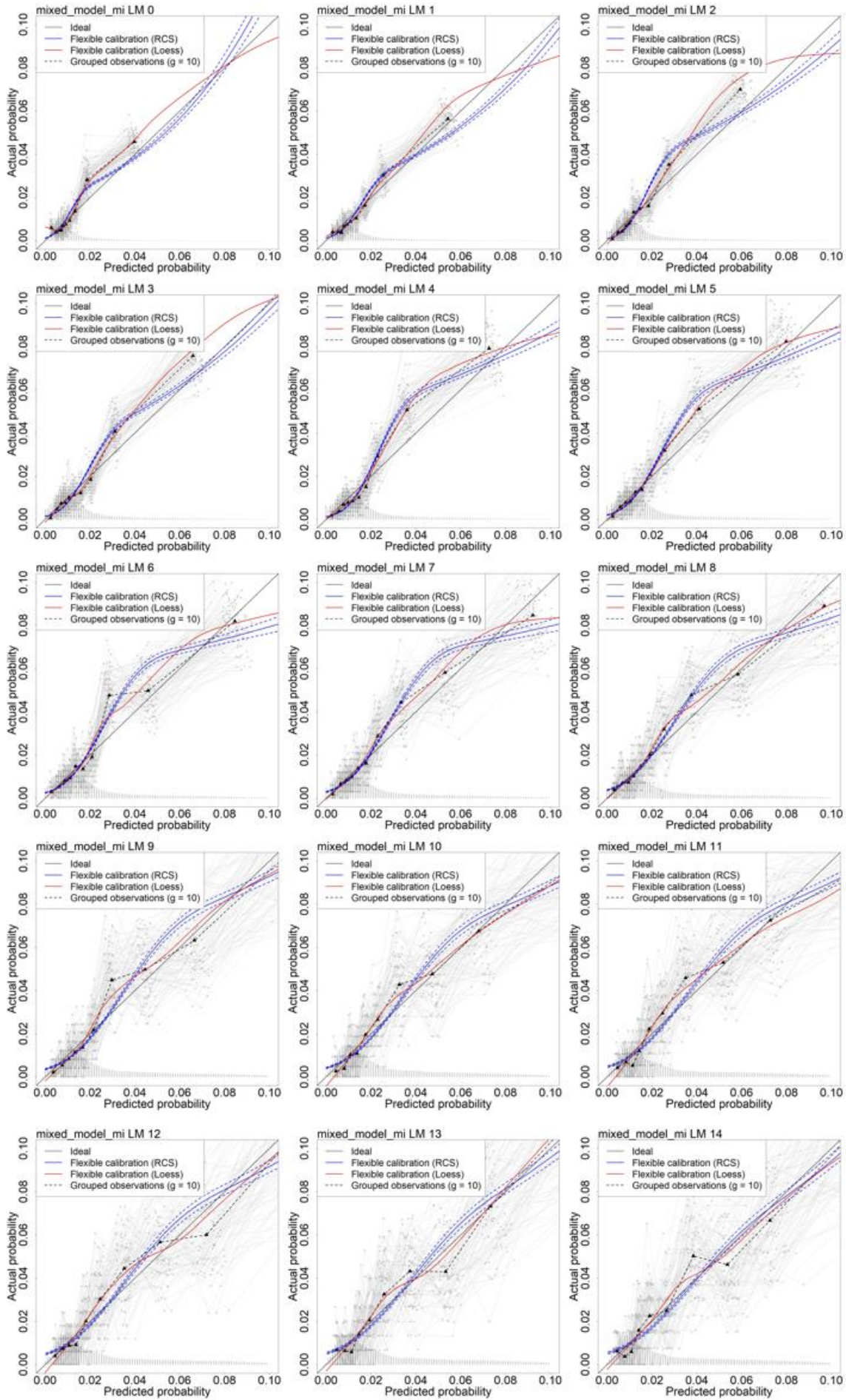

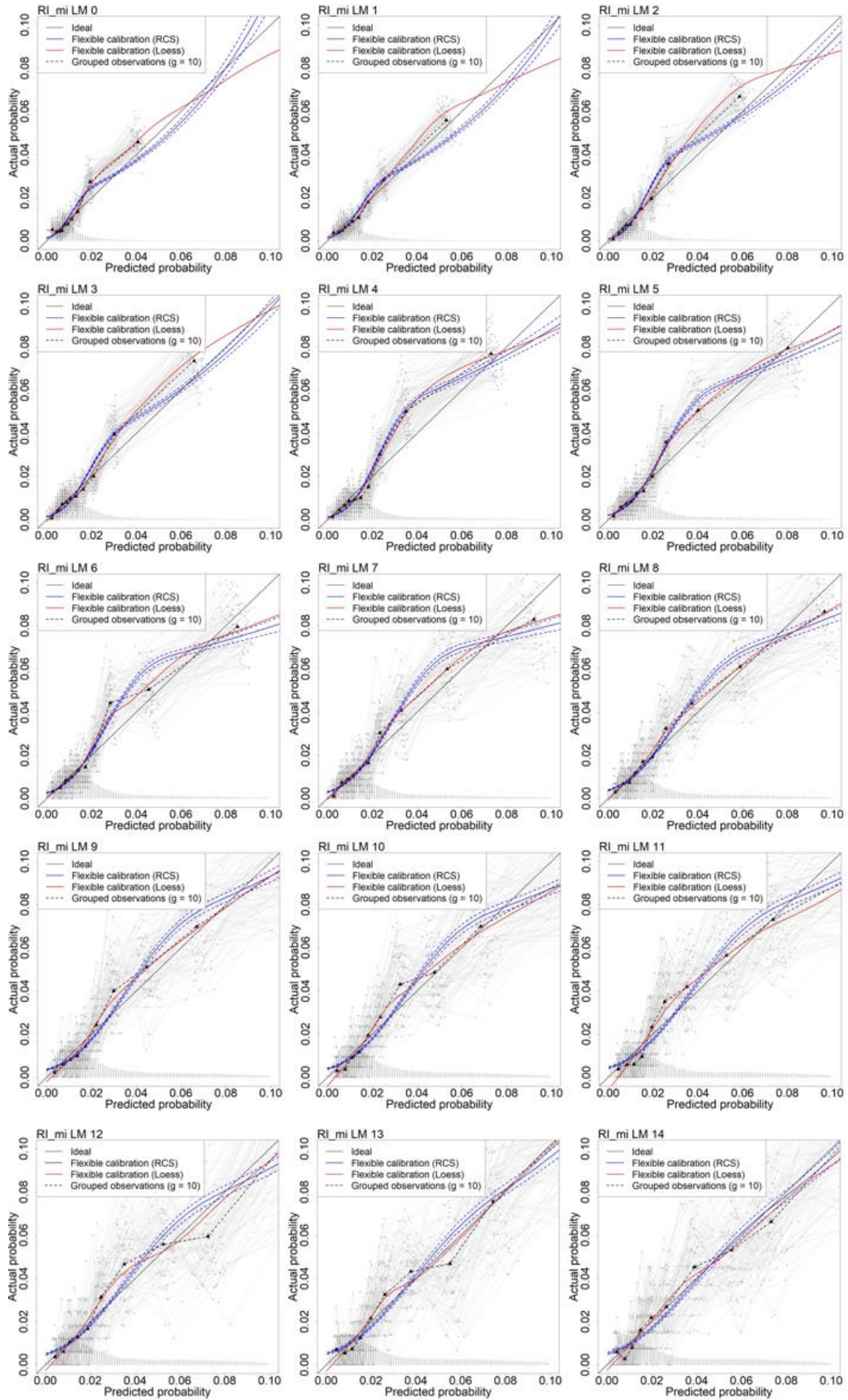

# Supplementary File 8: Comparison of all MICE methods

Figure S5 Comparison of performance metrics with MICE-xx, MICE-yx, MICE-yy

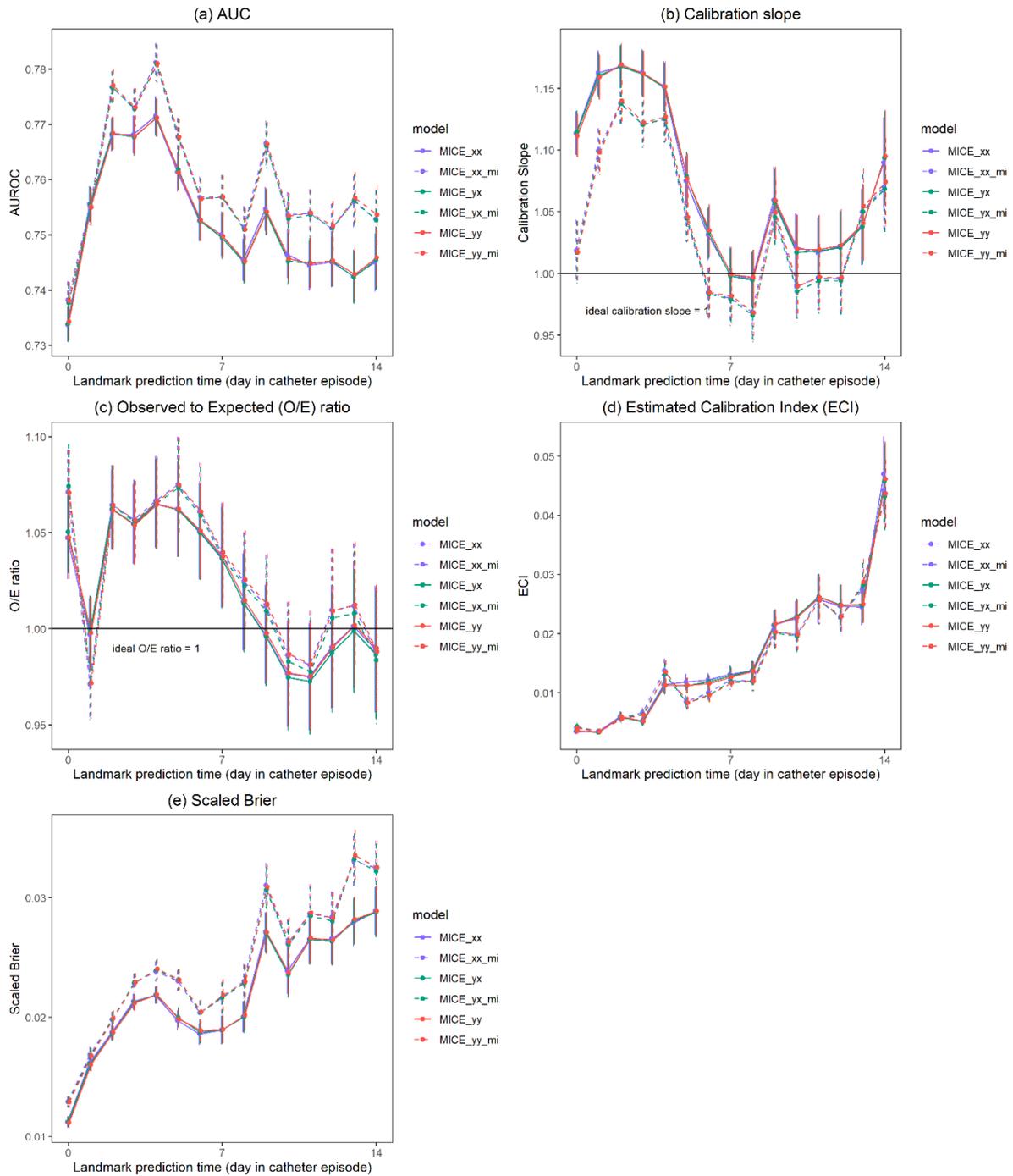

# Supplementary File 9: Runtimes for all imputation methods

Table S3 Runtimes for all imputation methods (unit: seconds)

| Type | Model | Median (IQR) | Range (min, max) | Model | Median (IQR) | Range (min, max) |
|---|---|---|---|---|---|---|
| impute | median_mode | 1.27 (1.25, 1.29) | (1.22, 1.73) | median_mode_mi | 1.31 (1.29, 1.35) | (1.26, 1.63) |
| impute | LOCF | 3.87 (3.8, 4.03) | (3.68, 4.53) | LOCF_mi | 5.96 (5.83, 6.06) | (5.63, 6.71) |
| impute | RI | 57.01 (56.5, 57.55) | (55.05, 63.44) | RI_mi | 71.35 (70.81, 71.99) | (69.54, 77.23) |
| impute | MICE_xx | 1790.64 (1778.7, 1800.78) | (1751.56, 1899.73) | MICE_xx_mi | 2487.52 (2476.58, 2501.56) | (2425.91, 2545.25) |
| impute | MICE_yx | 1841.06 (1823.96, 1853.88) | (1785.07, 1949.55) | MICE_yx_mi | 2486.65 (2472.35, 2500.38) | (2427.03, 2602.1) |
| impute | missForestPredict | 840.81 (667.35, 1006.08) | (457.26, 2307.97) | missForestPredict_mi | 1161.13 (903.93, 1468.42) | (582.32, 2893.69) |
| impute | mixed_model | 618.47 (601.93, 647.38) | (573.65, 713.14) | mixed_model_mi | 733.65 (714.32, 753.41) | (644.83, 844.09) |
| impute | missing indicator | 0.10 (0.1, 0.13) | (0.1, 0.44) | | | |
| build model | median_mode | 8.72 (8.6, 8.88) | (8.21, 18.3) | median_mode_mi | 13.71 (13.3, 15.17) | (12.68, 20.45) |
| build model | LOCF | 8.34 (8.17, 8.54) | (7.84, 16.18) | LOCF_mi | 13.18 (12.71, 14.59) | (11.7, 19.91) |
| build model | RI | 9.96 (9.82, 10.13) | (9.58, 16.72) | RI_mi | 14.54 (14.21, 15.96) | (13.54, 20.23) |
| build model | MICE_xx | 111.917 (111.38, 113.09) | (108.9, 123.22) | MICE_xx_mi | 173.40 (169.21, 186.82) | (162.86, 196.35) |
| build model | MICE_yx | 108.76 (107.56, 110.13) | (104.07, 114.91) | MICE_yx_mi | 164.88 (161.53, 177.69) | (154.97, 188.04) |
| build model | missForestPredict | 8.18 (7.82, 8.93) | (7.35, 95.66) | missForestPredict_mi | 16.76 (16.1, 17.91) | (14.89, 174.03) |
| build model | mixed_model | 8.68 (8.54, 8.83) | (8.2, 14.38) | mixed_model_mi | 13.38 (12.97, 14.67) | (12.29, 15.77) |
| build model | missing indicator | 13.93 (13.49, 15.21) | (12.9, 19.78) | | | |
| predict | median_mode | 82.72 (81.96, 83.93) | (80.32, 88.02) | median_mode_mi | 83.23 (82.51, 83.92) | (81.3, 89.13) |
| predict | LOCF | 83.16 (81.94, 84.64) | (80.28, 91.86) | LOCF_mi | 84.38 (83.7, 85.16) | (81.76, 90.52) |
| predict | RI | 83.29 (82.55, 84.16) | (80.6, 89) | RI_mi | 83.22 (82.5, 83.96) | (81.26, 90.16) |
| predict | MICE_xx | 796.00 (788.38, 804.16) | (770.62, 830.36) | MICE_xx_mi | 798.08 (790.8, 807.5) | (775.07, 832.73) |
| predict | MICE_yx | 812.41 (807.45, 817.48) | (786.18, 840.54) | MICE_yx_mi | 802.75 (795.56, 810.7) | (780.2, 833.81) |
| predict | missForestPredict | 145.71 (114.26, 183.69) | (87.95, 485.41) | missForestPredict_mi | 198.53 (160.53, 306.9) | (117.27, 515.2) |
| predict | mixed_model | 83.85 (83.09, 84.64) | (81.45, 94.76) | mixed_model_mi | 84.85 (84.06, 85.69) | (82.42, 88.62) |
| predict | missing indicator | 84.32 (83.55, 85.17) | (82.08, 91.81) | | | |